\documentclass[a4paper,11pt]{article}
\pdfoutput=1 

\usepackage{jheppub} 

\usepackage[T1]{fontenc} 
\usepackage{tensor}
\usepackage{subcaption}
\usepackage{mathtools}
\usepackage{dsfont}
\usepackage{xcolor}

\newcommand{\be}{\begin{equation}}
\newcommand{\ee}{\end{equation}}
\newcommand{\bea}{\begin{eqnarray}}
\newcommand{\eea}{\end{eqnarray}}
\newcommand{\beas}{\begin{eqnarray*}}
\newcommand{\eeas}{\end{eqnarray*}}
\newcommand{\ba}{\begin{array}}
\newcommand{\ea}{\end{array}}

\renewcommand{\Re}{\operatorname{Re}}

\newcommand{\hyperF}{{\tensor[_2]{F}{_1}}}
\newcommand{\AdS}{{\mathrm{AdS}}}
\newcommand{\gbl}{{\mathrm{g}}}
\newcommand{\Pcr}{{\mathrm{P}}}
\newcommand{\Sch}{{\mathrm{S}}}
\newcommand{\hrz}{+}

\DeclareMathOperator{\diag}{diag}
\DeclareMathOperator{\sech}{sech}

\DeclareMathOperator{\Vol}{Vol}

\title{\boldmath Holographic CFT states for localized perturbations to AdS black holes}

\author[a,b]{Hong Zhe Chen,}
\author[a]{Mark Van Raamsdonk,}

\affiliation[a]{
Department of Physics and Astronomy, University of British Columbia,\\
6224 Agricultural Road, Vancouver, B.C., V6T 1W9, Canada
}
\affiliation[b]{Perimeter Institute for Theoretical Physics, \\31 Caroline Street North, Waterloo, ON N2L 2Y5, Canada}

\emailAdd{hchen2@perimeterinstitute.ca}
\emailAdd{mav@phas.ubc.ca}

\abstract{In this note, we describe a holographic CFT construction of states dual to scalar perturbations of the maximally extended three-dimensional AdS-Schwarzschild black hole. The states are constructed by adding sources for a scalar operator to the path integral that constructs the thermofield double state. For a scalar field of arbitrary mass, we provide the general map between sources and scalar field perturbations at linear order. With this, we investigate to what extent it is possible using this construction to produce perturbations localized to one side of the black hole horizon. Our results suggest that it is possible to produce arbitrarily localized perturbations (and thus, plausibly, general linear perturbations to the black hole initial data slice), but that the amplitude of the perturbation must be taken small as the perturbation becomes more localized in order that the sources do not diverge, as found for the pure AdS case considered in arXiv:1709.10101.}

\begin{document}
\maketitle
\flushbottom

\section{Introduction}

The Euclidean path integral provides a powerful tool to construct states of holographic conformal field theories for which the corresponding state in the dual gravitational system has a nice classical description. Starting from the standard Euclidean path integral for the vacuum state,\footnote{Here, we take the path integral as being defined over $S \times I$, where $S$ is the spatial geometry on which the CFT lives and $I$ is the half-line parameterized by Euclidean time $\tau \in (- \infty,0]$. For the case where $S$ is a sphere, we could alternatively perform a conformal transformation to compactify this space to a ball.}
\begin{equation}
   \langle \phi_0 | \Psi \rangle =  \int^{\phi(\tau = 0) = \phi_0}_{\tau < 0} [d \phi] e^{-S_{Euc}}
\end{equation}
we can perturb the Euclidean action by sources for operators dual to the light fields in the bulk.
\begin{equation}
\label{sources}
    S_{Euc} \to S_{Euc} + \int dx d\tau \lambda_\alpha(x,\tau) {\cal O}_\alpha(x,\tau)
\end{equation}
If these sources vanish sufficiently rapidly for $\tau \to 0$, we define a perturbed state of the original theory.

The Lorentzian geometries dual to these states can be deduced making use of the real-time AdS/CFT formalism \cite{Marolf:2004fy,Skenderis:2008dh, Skenderis:2008dg}; see \cite{Botta-Cantcheff:2015sav, Christodoulou:2016nej} for early discussions and \cite{Botta-Cantcheff:2017qir,Marolf:2017kvq}. In \cite{Marolf:2017kvq}, the general map between sources and bulk perturbations was worked out explicitly at linear order for scalar and metric perturbations to Poincar\'e-AdS.

In this note, we make use of the same techniques to construct states dual to perturbations of AdS black hole geometries, focusing on scalar field perturbations of the AdS${}_3$ black holes for simplicity. Here, the starting point is the Euclidean path integral which constructs the thermofield double state of two copies of a CFT, namely the path integral on a cylinder $S^1 \times [-\beta/2,0]$. We consider sources which vanish near both ends of the cylinder and construct the map between the sources and the corresponding bulk perturbations at linear order.

An interesting aspect of this construction is that the sources necessarily affect the density matrix for both CFTs.\footnote{This is in contrast to the situation where we perturb the thermofield double state by acting with a unitary operator on one side.} Thus, we expect that the bulk perturbations generally affect both sides of the two-sided black hole geometry. As a key focus of this work, we investigate to what extent it is possible in this construction to localize perturbations to one side of the black hole, and to understand what types of Euclidean sources would give rise to such localization.

Making use of variational techniques, we numerically investigate the sources that optimize various measures of localization for the perturbations. Our results are consistent with the conclusion that it is possible by a careful choice of sources to produce perturbations that are arbitrarily well-localized to one side of the black hole or the other, and that the variance of these perturbation about a chosen point can also be made arbitrarily small. In this case, by taking linear combinations of sources that lead to localized perturbations, we should be able to choose sources which give rise to arbitrary initial data at the linearized level.

An interesting qualitative feature of our results is that the sources required to produce a perturbation of small variance that is well-localized to one side of the black hole are not well-localized on the corresponding side of the cylinder on which the path integral is defined. 
Instead, the required sources have a profile that is concentrated in the middle of the integration region (see figure \ref{fig:lambda_localizingPhi}). Another interesting qualitative feature, observed already in \cite{Marolf:2017kvq} for perturbations to pure AdS, is that decreasing the variance of fixed amplitude perturbations requires increasing the amplitude of the sources. Thus, to ensure validity of perturbation theory in the sources, the amplitude of bulk perturbations must be taken increasingly small for increasingly small variance.

We now provide a brief outline of the remainder of the paper. In section 2, we start by describing our basic setup for defining states using the Euclidean path integral. Next, we review the solution of the linearized scalar field equation of motion on Euclidean and Lorentzian AdS${}_3$ black hole backgrounds and derive the explicit relation between the Euclidean sources and Lorentzian initial data for scalar field perturbations. In section 3, we perform our numerical investigations to find sources which optimize various measures of localization to one side of the black hole for the perturbations.

The recent paper \cite{Botta-Cantcheff:2019apr} that appeared while this manuscript was in preparation also considers CFT states dual to perturbed black holes defined using path integral techniques and provides an interesting complementary discussion. While there is some overlap with our review of scalar field solutions on AdS${}_3$ and with the general construction, our main investigations of how to produce localized perturbations does not overlap with the contents of \cite{Botta-Cantcheff:2019apr}.

\section{Linearized black hole perturbations from path-integral sources}

\begin{figure}
    \centering
    \includegraphics[width=0.7\textwidth]{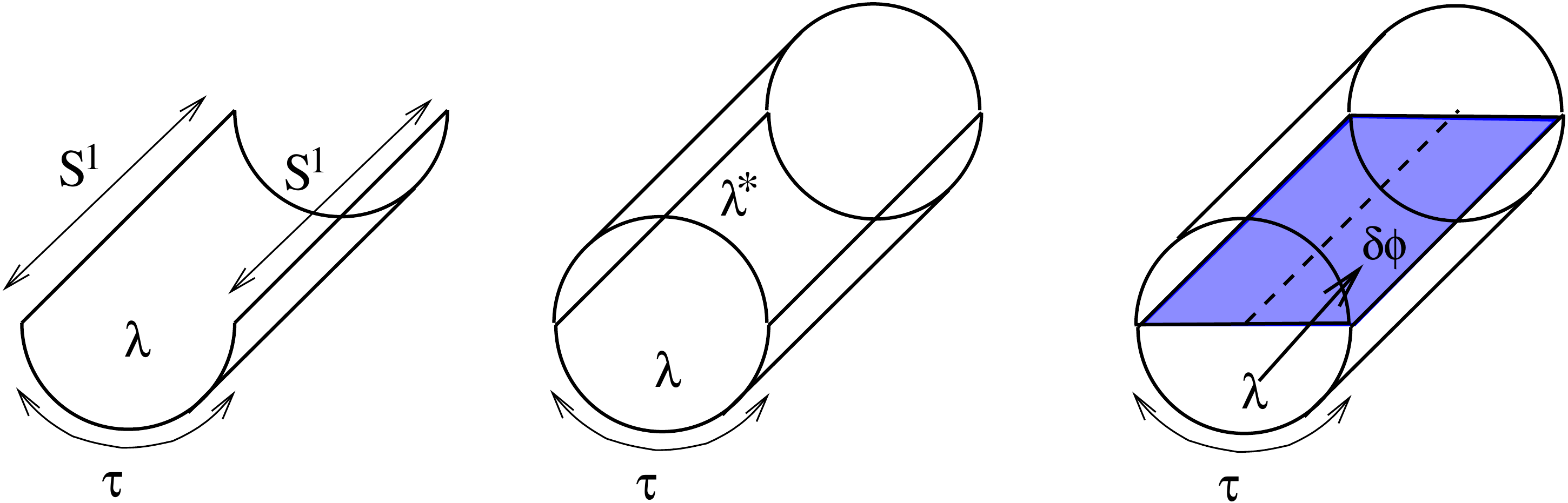}
  \caption{Left: path integral for a perturbed thermofield double state of two CFTs on $S^1$ (ends are periodically identified). Middle: Path integral used to compute $t=0$ observables for this state. Right: Euclidean gravity solution dual to the perturbed CFT state (interior of torus. Initial data for the Lorenzian geometry corresponding to our state is obtained via analytic continuation of the fields on the blue surface.}
  \label{BHpert}
\end{figure}

In this paper, we focus on scalar perturbations to the two-sided AdS${}_3$ black hole geometry. In the CFT description, this spacetime corresponds to the thermofield double state of a pair of CFTs each on a spatial circle. The state can be prepared by a path integral on a cylinder $S^1 \times [-\beta/2,0]$:
\begin{equation}
    \langle \phi_L \phi_R | \Psi \rangle = \int_{\phi(-\beta/2) = \phi_L, }^{\phi(0) = \phi_R, } [d \phi(\tau,x)] e^{-S_{Euc}} \; .
\end{equation}
Here, $\phi$ represents the full set of CFT fields. To produce perturbed black hole states, we can perturb the Euclidean action here by sources as in (\ref{sources}). At the linearized level, if we wish to introduce perturbations to a particular field in the gravitational theory, we can add a source for the corresponding operator. In this paper, we focus on scalar field perturbations.

To understand the geometries dual to these perturbed states, we apply the basic recipe of \cite{Skenderis:2008dg}, reviewed in detail in \cite{Marolf:2017kvq} and displayed in figure \ref{BHpert}. We consider a CFT path-integral defined on $S^1 \times [-\beta/2,\beta/2]$ with identification $\beta/2 \equiv -\beta/2$ and sources for $\tau > 0$ defined by $\lambda(\tau, x) = \lambda^*(-\tau,x)$. Via the standard AdS/CFT dictionary, we can associate to this path integral a corresponding Euclidean gravity configuration which solves the gravitational equations subject to the boundary conditions that the boundary metric is the torus geometry on which the path integral is defined, and the asymptotic values of the fields are determined by the sources we add. In the unperturbed geometry, the bulk slice that divides the spacetime symmetrically and asymptotes to the $\tau = 0$ and $\tau = \pm \beta/2$ circles at the boundary gives the initial data for the corresponding Lorentzian geometry. This includes two asymptotic regions connected by an Einstein-Rosen bridge.

Working perturbatively, the same slice of the bulk geometry will correspond to the initial data for the Lorentzian solution, and the Lorentzian perturbations are determined directly from the scalar field perturbations on this surface produced by the sources in the Euclidean solution via\footnote{At higher orders in perturbation theory or non-perturbatively, the relation between the Euclidean and Lorentzian perturbations is more complicated. However,  for cases with real sources, the resulting Euclidean spacetime will have a time-reflection symmetry and the Lorentzian initial data (which will have vanishing time derivatives for the fields) can be read off directly from the spatial slice lying at the fixed point of this symmetry. }
\beas
    \delta \phi_L(x,t=0) &=& \delta \phi_E(x,\tau=0) \cr
    \partial_t \delta \phi_L(x,t=0) &=& i\partial_\tau \delta \phi_E(x,\tau=0) \; .
\eeas
At the linearized level, the Euclidean perturbations on the right side here are determined by the sources via a Euclidean boundary-to-bulk propagator.

Without sources, and for $\beta < R_{S^1}$, the gravitational configuration is just global Euclidean AdS with a periodic identification of the usual Euclidean time direction, which is reinterpreted as the direction corresponding to the spatial coordinate of the CFT. In the next subsection, we recall the scalar field solutions in the standard global AdS coordinates and then make the reinterpretation to obtain the desired solutions for the Euclidean black hole background.

\subsection{Scalar field solution in Euclidean global $\AdS_3$}
\label{eq:scalarFieldSolution_EuclideanGlobalAdS}
In this section, we recall the classical scalar field solution in Euclidean global $\AdS_3$. This will be used in the next section where we reinterpret periodically identified global $\AdS_3$ as a Euclidean black hole.

Euclidean global $\AdS$ can be described using the metric
\begin{align*}
  ds^2
  =& \frac{\ell^2}{\cos^2(\rho_\gbl)}[d\tau_\gbl^2+d\rho_\gbl^2+\sin^2(\rho_\gbl)d\theta_\gbl^2].
\end{align*}
The classical equation of motion for a scalar field of mass $\mu$ in this geometry is
\begin{align*}
  0
  =& (g^{\mu\nu}\nabla_\mu \nabla_\nu-\mu^2)\Phi
     = \frac{1}{\sqrt{g}}\partial_\mu\left(\sqrt{g}g^{\mu\nu}\partial_\nu\Phi\right)-\mu^2\Phi
  \\
  =& \frac{1}{\ell^2}\left[
     \cos^2(\rho_\gbl)(\partial_{\tau_\gbl}^2+\partial_{\rho_\gbl}^2)
     +\cot(\rho_\gbl)\partial_{\rho_\gbl}
     +\cot^2(\rho_\gbl)\partial_{\theta_\gbl}^2
     \right]\Phi
     -\mu^2\Phi.
\end{align*}
Via separation of variables, we can expand the solutions in terms of mode functions
\begin{equation}
e^{i(\omega_\gbl\tau_\gbl + m_\gbl\theta_\gbl)} R_{m_\gbl}(\omega_\gbl,\rho_\gbl) \; .
\end{equation}
where the radial function $R_{m_\gbl}$ satisfies a second order differential equation in $\rho_\gbl$,
\begin{align}
  0
  =& \frac{1}{\ell^2}\left[
     \cos^2(\rho_\gbl)(-\omega_\gbl^2+\partial_{\rho_\gbl}^2)
     +\cot(\rho_\gbl)\partial_{\rho_\gbl}
     -m_\gbl^2 \cot^2(\rho_\gbl)
     \right] R_{m_\gbl}(\omega_\gbl,\rho_\gbl)
     -\mu^2 R_{m_\gbl}(\omega_\gbl,\rho_\gbl).
     \label{eq:EqnOfMotion_EuclideanGlobalAdS_rho}
\end{align}
We shall focus on solutions which do not diverge in the bulk. If we rescale the field
\begin{align}
  R_{m_\gbl}(\omega_\gbl,\rho_\gbl)
  \equiv& \cos^{1+\nu}(\rho_\gbl)\sin^{|m_\gbl|}(\rho_\gbl)
          \tilde{R}_{m_\gbl}(\omega_\gbl,\sin^2 \rho_\gbl), \notag\\
  \nu
  \equiv& \sqrt{1+\ell^2 \mu^2}
          = \Delta_{\cal O}-1,
          \label{eq:nu}
\end{align}
then \eqref{eq:EqnOfMotion_EuclideanGlobalAdS_rho} becomes a hypergeometric differential equation \eqref{eq:hyperDE} in the variable $\sin^2\rho_\gbl$, with
\begin{align*}
  a =& \frac{1+\nu +|m_\gbl|-i\omega_\gbl}{2},
  & b=& \frac{1+\nu + |m_\gbl|+i\omega_\gbl}{2},
  & c=& 1 + |m_\gbl|.
\end{align*}
Taking $ \tilde{R}_{m_\gbl}(\omega_\gbl,\sin^2 \rho_\gbl)$ to be a solution of the form \eqref{eq:hyperDE_z0_sol1}, re-expressed using \eqref{eq:hyperF_equivalent_form3}, we find
\begin{align}
  R_{m_\gbl}(\omega_\gbl,\rho_\gbl)
  =& a_{m_\gbl}(\omega_\gbl) \cos^{1-\nu}(\rho_\gbl) \sin^{|m_\gbl|}(\rho_\gbl) \notag\\
     &\times \hyperF\left(
     \frac{1-\nu+|m_\gbl|-i\omega_\gbl}{2},
     \frac{1-\nu+|m_\gbl|+i\omega_\gbl}{2};
     1+|m_\gbl|;
     \sin^2(\rho_\gbl)
       \right),
       \label{eq:Phi_EuclideanGlobalAdS}
\end{align}
where we choose
\begin{align}
  a_{m_\gbl}(\omega)
    \equiv& \frac{
            \left|\Gamma\left(\frac{1+\nu+|m_\gbl|-i\omega_\gbl}{2}\right)\right|^2
            }{
            \Gamma(|m_\gbl|+1)\Gamma(\nu)
            }.
     \label{eq:C_EuclideanGlobalAdS}
\end{align}
so that (using the identity \eqref{eq:hyperF_z1})
\be
\label{Rasympt}
R_{m_\gbl}(\omega_\gbl,\rho_\gbl) \sim \cos^{1 - \nu} (\rho_g)
\ee
for $\rho_g \to \pi/2$. When writing \eqref{eq:Phi_EuclideanGlobalAdS}, we have chosen to use the alternative form \eqref{eq:hyperF_equivalent_form3} of the hypergeometric function in order to emphasize the $\sim \cos^{1-\nu}(\rho_\gbl)$ behaviour of $\Phi$ near the boundary $\rho_\gbl=\pi/2$.

For our problem, we would like to find solutions where the asymptotic behavior is related to the source function $\lambda(\tau_g,\theta_g)$ for the associated scalar operator in the CFT by the usual holographic dictionary,
\begin{align*}
  \lim_{\rho_\gbl \to \pi/2} \frac{\Phi(\tau_\gbl,\rho_\gbl,\theta_\gbl)}{\cos^{1-\nu}(\rho_\gbl)}
  =& \lambda(\tau_\gbl,\theta_\gbl) \; .
\end{align*}
Writing
\begin{align*}
  \lambda(\tau_\gbl,\theta_\gbl)
  =& \int \frac{d\omega_\gbl}{2\pi} \sum_{m_\gbl} e^{i(\omega_\gbl\tau_\gbl + m_\gbl\theta_\gbl)}\lambda_{m_\gbl}(\omega_\gbl),
\end{align*}
the correct linear combination of mode functions is
\begin{align}
  \Phi(\tau_\gbl,\rho_\gbl,\theta_\gbl)
  =& \int \frac{d\omega_\gbl}{2\pi} \sum_{m_\gbl} \lambda_{m_\gbl}(\omega_\gbl) e^{i(\omega_\gbl\tau_\gbl + m_\gbl\theta_\gbl)} R_{m_\gbl}(\omega_\gbl,\rho_\gbl) \; .
     \label{eq:Phi_modeExpansion_EuclideanGlobalAdS}
\end{align}

This gives the linear map between sources and bulk scalar perturbations when the CFT is on an infinite cylinder and the bulk geometry is Euclidean global AdS.

\subsection{From sources to perturbations for the Euclidean black hole}
\label{sec:initial_data}
\begin{figure}
  \centering
  \includegraphics[width=0.3\textwidth]{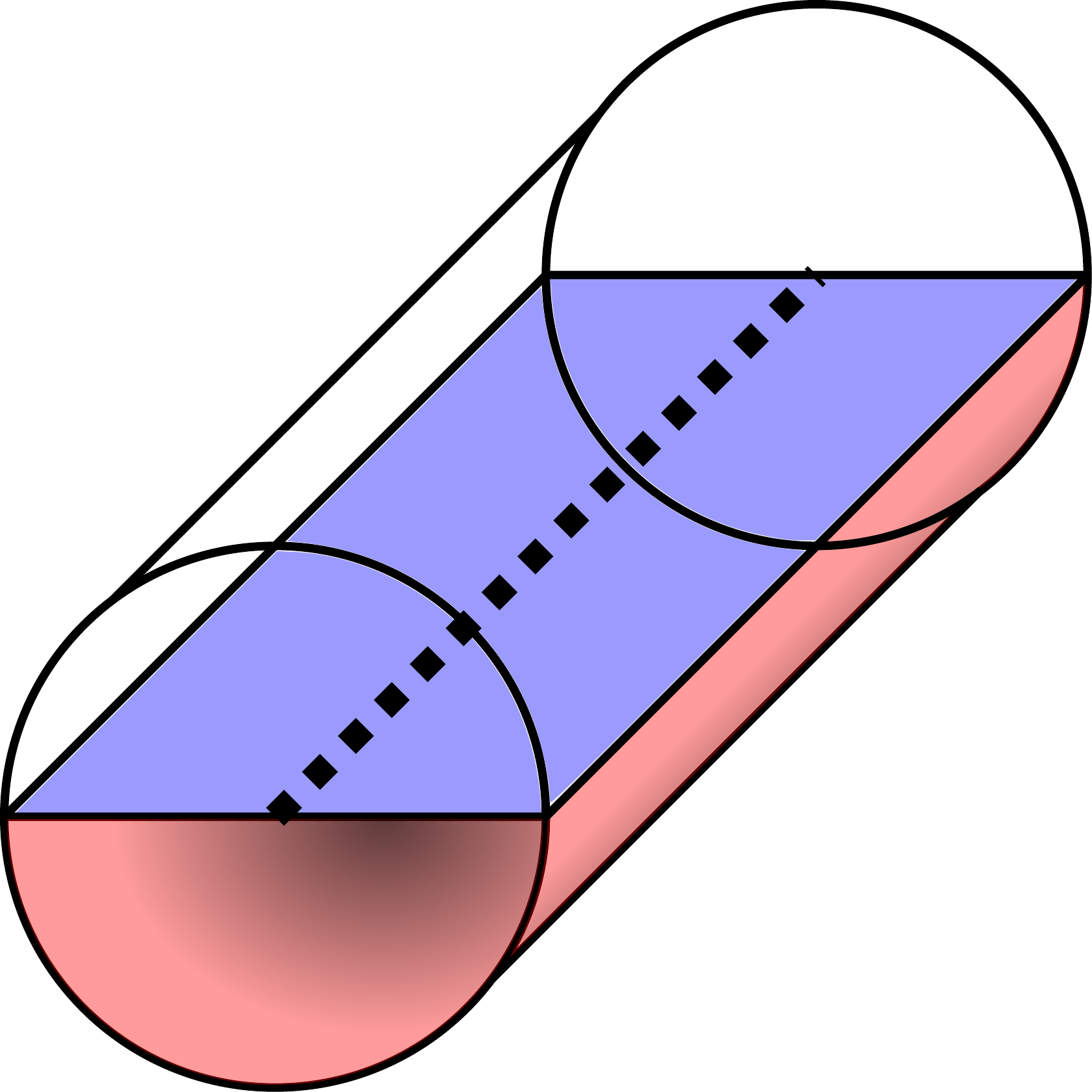}
  \caption{To obtain a perturbed thermofield double state, sources are inserted into the Euclidean path integral over the surface shown in red for global $\AdS_3$. The initial data are then read off from the blue surface. The dotted line indicates the horizon dividing the two black hole exteriors. If the two disks at the ends of the global $\AdS_3$ cylinder
    are identified, then the black hole is spherical. A planar black hole is obtained by making the global $\AdS_3$ cylinder infinitely long. }
  \label{fig:AdS_sourcesAndInitialData}
\end{figure}

\begin{figure}
  \centering
  \begin{subfigure}[c]{0.49\textwidth}
    \includegraphics[width=\textwidth]{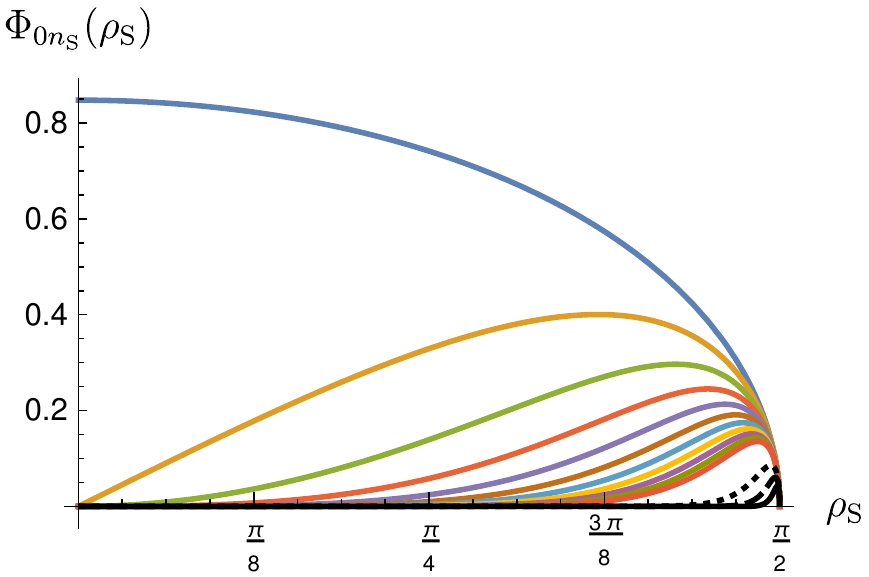}
    \caption{$\Phi_{0n_\Sch}$ as a function of $\rho_\Sch$.}
    \label{fig:Phi_0n_rho}
  \end{subfigure}
  \begin{subfigure}[c]{0.49\textwidth}
    \includegraphics[width=\textwidth]{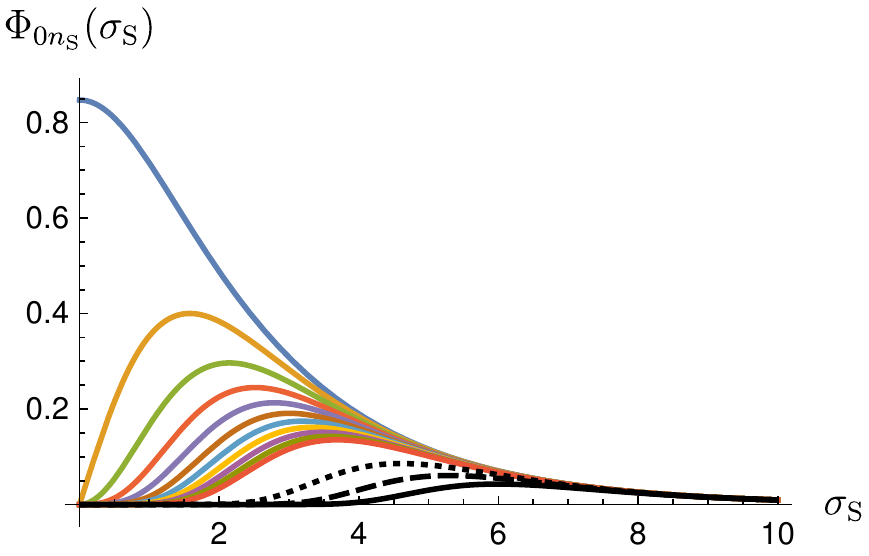}
    \caption{$\Phi_{0n_\Sch}$ as a function of $\sigma_\Sch$.}
    \label{fig:Phi_0n_sigma}
  \end{subfigure}
  \caption{$\Phi_{0n_\Sch}$ plotted against $\rho_\Sch$ and $\sigma_\Sch$ for $n_\Sch=0,1,\ldots,10$ (coloured) and $n_\Sch=25,50,100$ (black). The upper-most and bottom-most coloured curves are for $n_\Sch=0,10$ respectively and the black dotted, dashed, and solid curves are for $n_\Sch=25,50,100$ respectively. Here, $\nu=1/2$.}
  \label{fig:Phi_0n}
\end{figure}

As described in \S\ref{sec:sidewaysGlobalAdS_asBlackHole}, the two-sided AdS-black hole (i.e. BTZ) geometry can be described by coordinates $(\tau_\Sch,\rho_\Sch,\theta_\Sch)$ with the metric \eqref{eq:SchwarzschildAdS_EuclideanMetric} of a global $\AdS_3$ cylinder lying sideways. Here, $\tau_\Sch$ is the Euclidean analytic continuation of the usual Schwarzschild black hole time $t_\Sch$;  $\rho_\Sch$ is related to the more familiar Schwarzschild radial coordinate $r_\Sch$ by \eqref{eq:SchwarzschildAdS_rho_r}; and $\theta_\Sch$ is the usual Schwarzschild angular coordinate. Since the metric \eqref{eq:SchwarzschildAdS_EuclideanMetric} is obtained from the usual global $\AdS_3$ metric \eqref{eq:globalAdS_metric_rho} with the replacements \eqref{eq:globalAdS_to_SchwarzschildAdS_coords} (where the black hole radius $r_\hrz$ is related to temperature $1/\beta$ by \eqref{eq:blackHole_temperature_radius}), we just need to apply the same replacements to our results from the previous section in order to study thermofield double/BTZ black hole correspondence. We will also rename:
\begin{align}
  m_\gbl \to& n_\Sch, & \omega_\gbl\to k_{m_\Sch} \equiv \frac{\beta m_\Sch}{2\pi \ell}, & &(m_\Sch,n_\Sch \in \mathbb{Z})
                                                                                             \label{eq:globalAdS_to_SchwarzschildAdS_freqs}
\end{align}
The latter discretizes the frequencies allowed in $\theta_\Sch$ so that the BTZ identification \eqref{eq:BTZ_identification_Schwarzschild} is satisfied. Making these replacements in \eqref{eq:Phi_EuclideanGlobalAdS} and \eqref{eq:C_EuclideanGlobalAdS}, we find that the classical scalar field solution in the Euclidean BTZ black hole is
\begin{align}
  \Phi(\tau_\Sch,\rho_\Sch,\theta_\Sch)
  =& \sum_{m_\Sch,n_\Sch} \lambda_{m_\Sch n_\Sch} e^{i\left(\frac{2\pi n_\Sch \tau_\Sch}{\beta} + m_\Sch \theta_\Sch\right)}
     \Phi_{m_\Sch n_\Sch}(\rho_\Sch),
     \label{eq:Phi}
\end{align}
where
\begin{align}
  \Phi_{m_\Sch n_\Sch}(\rho_\Sch)
  =& a_{m_\Sch n_\Sch} \cos^{1-\nu}(\rho_\Sch) \sin^{|n_\Sch|}(\rho_\Sch) \notag\\
     &\times \hyperF\left(
     \frac{1-\nu+|n_\Sch|-ik_{m_\Sch}}{2},
     \frac{1-\nu+|n_\Sch|+ik_{m_\Sch}}{2};
     1+|n_\Sch|;
     \sin^2(\rho_\Sch)
       \right) \label{eq:Phi_mn}
  \\
  a_{m_\Sch n_\Sch}
  =& \frac{
     \left|\Gamma\left(\frac{1+\nu+|n_\Sch|-ik_{m_\Sch}}{2}\right)\right|^2
     }{
     \Gamma(|n_\Sch|+1)\Gamma(\nu)
     }
  \notag\\
  \lambda(\tau_\Sch,\theta_\Sch)
  =& \sum_{m_\Sch,n_\Sch} e^{i\left(\frac{2\pi n_\Sch \tau_\Sch}{\beta} + m_\Sch \theta_\Sch\right)}
     \lambda_{m_\Sch n_\Sch}.
     \notag
\end{align}
The radial dependence of the various $m_\Sch=0$ modes $\Phi_{0 n_\Sch}$ of the field are plotted in Figure \ref{fig:Phi_0n_rho} as functions of $\rho_\Sch$. In Figure \ref{fig:Phi_0n_sigma}, we also plot $\Phi_{0 n_\Sch}$ as a function of an alternative radial coordinate $\sigma_\Sch$, given by \eqref{eq:SchwarzschildAdS_rho_sigma}, which will be useful later.

The initial data for the two sides of the black hole can be read off from \eqref{eq:Phi} on the slices at $\tau_\Sch=0$ and $\tau_\Sch=-\beta/2$:
\begin{align}
  \Phi^{(1)}(t_\Sch=0,\rho_\Sch,\theta_\Sch)
  =& \Phi(\tau_\Sch=0,\rho_\Sch,\theta_\Sch)
  = \sum_{m_\Sch,n_\Sch} \lambda_{m_\Sch n_\Sch} e^{im_\Sch \theta_\Sch}
     \Phi_{m_\Sch n_\Sch} (\rho_\Sch)
     \label{eq:initial_data_Phi1}
  \\
  \partial_{t_\Sch} \Phi^{(1)}(t_\Sch=0,\rho_\Sch,\theta_\Sch)
  =& i\partial_{\tau_\Sch} \Phi(\tau_\Sch=0,\rho_\Sch,\theta_\Sch)
  = -\sum_{m_\Sch,n_\Sch} \frac{2\pi n_\Sch}{\beta} \lambda_{m_\Sch n_\Sch} e^{i m_\Sch \theta_\Sch}
     \Phi_{m_\Sch n_\Sch} (\rho_\Sch)
     \label{eq:initial_data_dPhi1}
  \\
  \Phi^{(2)}(t_\Sch=0,\rho_\Sch,\theta_\Sch)
  =& \Phi(\tau_\Sch=-\beta/2,\rho_\Sch,\theta_\Sch)
  = \sum_{m_\Sch,n_\Sch} (-1)^{n_\Sch} \lambda_{m_\Sch n_\Sch} e^{i m_\Sch \theta_\Sch}
     \Phi_{m_\Sch n_\Sch} (\rho_\Sch)
     \label{eq:initial_data_Phi2}
  \\
  \partial_{t_\Sch} \Phi^{(2)}(t_\Sch=0,\rho_\Sch,\theta_\Sch)
  =& i\partial_{\tau_\Sch} \Phi(\tau_\Sch=-\beta/2,\rho_\Sch,\theta_\Sch) \\
  =& -\sum_{m_\Sch,n_\Sch} \frac{2\pi n_\Sch}{\beta} (-1)^{n_\Sch} \lambda_{m_\Sch n_\Sch} e^{im_\Sch \theta_\Sch}
     \Phi_{m_\Sch n_\Sch} (\rho_\Sch).
     \label{eq:initial_data_dPhi2}
\end{align}
We have taken the convention that time evolution is in the $-t_\Sch$ direction in the second exterior, with the future and past horizons at $t_\Sch=-\infty,+\infty$ respectively.

Note that
\begin{align}
  \Phi_{m_\Sch,n_\Sch}(\rho_\Sch)
  =& \Phi_{m_\Sch,-n_\Sch}(\rho_\Sch)
     \label{eq:Phi_mn_sign_of_n}
\end{align}
so \eqref{eq:initial_data_Phi1} and \eqref{eq:initial_data_Phi2} only depend on the sources through the combinations
\begin{align*}
  \lambda_{m_\Sch,0},\qquad \lambda_{m_\Sch,n_\Sch}+\lambda_{m_\Sch,-n_\Sch}
  \qquad(m_\Sch\in\mathbb{Z},n_\Sch\in\mathbb{N})
\end{align*}
and \eqref{eq:initial_data_dPhi1} and \eqref{eq:initial_data_dPhi2} only depend on the sources through the combinations
\begin{align*}
  \lambda_{m_\Sch,n_\Sch}-\lambda_{m_\Sch,-n_\Sch}
  \qquad(m_\Sch\in\mathbb{Z},n_\Sch\in\mathbb{N}).
\end{align*}

Finally, we remark on some contstraints placed on the sources. Since we must Hermitian conjugate to go from a ket to a bra, we have
\begin{align}
  \lambda(\tau_\Sch,\theta_\Sch) =& \lambda^\dagger(-\tau_\Sch,\theta_\Sch),
  & &\text{i.e.},
  & \lambda_{m_\Sch,n_\Sch}=\lambda_{-m_\Sch,n_\Sch}^\dagger.
    \label{eq:lambdaConstraint_timeReversal}
\end{align}
To have the sources vanish at $\tau_\Sch=0,\beta/2$, we require
\begin{align}
  0
  =& \sum_{\substack{m_\Sch,n_\Sch \\ n_\Sch\text{ even}}} e^{im_\Sch\theta_\Sch} \lambda_{m_\Sch n_\Sch}
  = \sum_{\substack{m_\Sch,n_\Sch \\ n_\Sch\text{ odd}}} e^{im_\Sch\theta_\Sch} \lambda_{m_\Sch n_\Sch}.
  \label{eq:lambdaConstraint_vanishAtInitialTime}
\end{align}
If we also assume that the sources $\lambda(\tau_\Sch,\theta_\Sch)$ are real, then
\begin{align}
  \lambda_{m_\Sch,n_\Sch} =& \lambda_{m_\Sch, -n_\Sch}.
                             \label{eq:lambdaConstraint_real}
\end{align}

\subsection{Special case: $\nu=1/2$}
As a check of our work, we now produce an equivalent form for classical scalar field solutions by starting in Poincar\'e coordinates and focusing on the special case
\begin{align*}
  \ell^2\mu^2 =& -\frac{3}{4},
  & \Delta =& \frac{3}{2},
  & \nu =& \frac{1}{2},
\end{align*}
as was done in \S 4 of \cite{Marolf:2017kvq}. Setting $\nu=1/2$ simplifies calculations because, in Poincar\'e coordinates \eqref{eq:Poincare_z}-\eqref{eq:Poincare_x}, the classical equation of motion for the rescaled field $\Phi/\sqrt{z_\Pcr}$ is Laplace's equation,
\begin{align*}
  0 =& (\partial_{\tau_\Pcr}^2 + \partial_{z_\Pcr}^2 + \partial_{x_\Pcr}^2) \frac{\Phi}{\sqrt{z_\Pcr}}.
\end{align*}
The general solution, in spherical Poincar\'e coordinates \eqref{eq:spherical_Poincare_coordinates} is
\begin{align*}
  \Phi(r_\Pcr,\theta_\Pcr,\varphi_\Pcr)
  =& \sum_{\alpha\in A} \sum_{n_\Pcr}
     \lambda_{\alpha n_\Pcr} b_{\alpha n_\Pcr}
     \left(\frac{r_\Pcr}{\ell}\right)^{\alpha+\frac{1}{2}} e^{in_\Pcr \varphi_\Pcr}
     \sqrt{\cos\theta_\Pcr} P_\alpha^{|n_\Pcr|}(\cos(\theta_\Pcr)),
\end{align*}
where the powers of $r_\Pcr$ in $\{r_\Pcr^\alpha : \alpha\in A\}$ form a basis for functions of $r_\Pcr$, $b_{\alpha n_\Pcr}$ are 
normalization constants to be determined by imposing the holographic dictionary, and $P_a^b$ is the Ferrer function of degree $a$ and order $b$ (see \S\ref{sec:LegendreFerrersP}). In terms of the Schwarzschild coordinates which we have been using previously in \S\ref{sec:initial_data},
\begin{align*}
  \Phi(\tau_\Sch,\rho_\Sch,\theta_\Sch)
  =& \sum_{\alpha\in A} \sum_{n_\Sch}
     \lambda_{\alpha n_\Sch} b_{\alpha n_\Sch}
     \exp\left[\left(\alpha+\frac{1}{2}\right)\frac{2\pi\ell }{\beta}\theta_\Sch + \frac{i2\pi n_\Sch }{\beta}\tau_\Sch\right]
     \sqrt{\cos\rho_\Sch} P_\alpha^{|n_\Sch|}(\cos(\rho_\Sch)).
\end{align*}
We should pick $\alpha$ so that $\Phi$ is periodic in $\theta_\Sch$, satisfying the BTZ identification \eqref{eq:BTZ_identification_Schwarzschild}. One choice is
\begin{align*}
  \alpha =& \left\{
       -\frac{1}{2} + ik_{m_\Sch} : m_\Sch\in\mathbb{Z}
       \right\},
\end{align*}
giving
\begin{align}
  \Phi(\tau_\Sch,\rho_\Sch,\theta_\Sch)
  =& \sum_{m_\Sch n_\Sch}
     \lambda_{m_\Sch n_\Sch} b_{m_\Sch n_\Sch}
     e^{i\left(m_\Sch \theta_\Sch + \frac{2\pi n_\Sch}{\beta} \tau_\Sch\right)}
     \sqrt{\cos\rho_\Sch} P_{-\frac{1}{2} + ik_{m_\Sch}}^{|n_\Sch|}(\cos(\rho_\Sch)).
     \label{eq:Phi_nu_half}
\end{align}
Using the value \eqref{eq:FerrersP_x0} of the Ferrer function at $\cos(\rho_\Sch)=0$, we have
\begin{align*}
  b_{m_\Sch n_\Sch}
  \equiv& \left(\frac{1}{2}\right)^{|n_\Sch|} \frac{1}{\sqrt{\pi}}
     \left|\Gamma\left(\frac{3}{4}-\frac{|n_\Sch|+ik_{m_\Sch}}{2}\right)\right|^2.
\end{align*}
It is easy to check, by applying \eqref{eq:hyperF_FerrersP} then \eqref{eq:LegendreP_degreeSignFlip} and \eqref{eq:LegendreP_orderSignFlip}, that \eqref{eq:Phi_mn} at $\nu=1/2$ equals $b_{m_\Sch n_\Sch}$ times the $\rho_\Sch$ dependent part of \eqref{eq:Phi_nu_half}.

\subsection{Comments on Lorentzian evolution}
Before moving on to numerically investigating the relationship between Euclidean sources and the localization of initial data, let us provide some brief comments on the evolution of the scalar field initial data in Lorentzian time. We refer the reader to \cite{Cardoso:2001hn,Balasubramanian:2004zu,Botta-Cantcheff:2019apr} for more detailed discussions.

To work out the Lorentzian evolution, we can follow the basic strategy used in \cite{Marolf:2017kvq} for perturbations to pure AdS; that is, we start with a basis of appropriately normalized solutions to the Lorentzian field equations and then choose the appropriate linear combination by matching to our initial data. With vanishing Lorentzian sources, the holographic dictionary requires normalizable solutions to vanish as $\sim \cos^{1+\nu}(\rho_\Sch)$ (in $3$-dimensional spacetime) at the boundary. When the appropriate Lorentzian solution is found, the asymptotic behavior of the scalar field near the AdS boundary gives the CFT one-point function for the CFT primary operator dual to the bulk field. 

We note that for real sources, the initial data has vanishing time-derivatives, so the Lorentzian solutions will be time-symmetric, with matter emerging from the past horizon of the black hole and falling in to the future horizon. 


\section{Sources for localized perturbations}

Here, we perform numerical calculations to examine various properties of the scalar field initial data. For our numerical calculations, we will stick with $\nu=1/2$. Additionally, for simplification, we will consider sources $\lambda_{0n_\Sch}$ and initial data $\Phi_{0n_\Sch}$ which are independent of the Schwarzschild angular coordinate $\theta_\Sch$. We shall work in the radial coordinate $\sigma_\Sch$, given by \eqref{eq:SchwarzschildAdS_rho_sigma}, since it is nicely related to proper length on constant $t_\Sch,\theta_\Sch$ lines by $ds^2 = \ell^2 d\sigma_\Sch^2$.

\subsection{Maximizing the ratio of $L_{\sigma_\Sch}^2$ norms of $\Phi^{(1)}(t_\Sch=0),\Phi^{(2)}(t_\Sch=0)$}
\label{sec:maximizingPhi1Phi2Ratio}
In this section, we wish to maximize the ratio
\begin{align*}
  N[\Phi^{(1)},\Phi^{(2)}]
  \equiv& \frac{L^2_{\sigma_\Sch}[\Phi^{(1)}(t_\Sch=0)]}{L^2_{\sigma_\Sch}[\Phi^{(2)}(t_\Sch=0)]},
  & L^2_{\sigma_\Sch}[f(\sigma_\Sch)]
  \equiv& \int d\sigma_\Sch \; |f(\sigma_\Sch)|^2,
\end{align*}
for sources and initial data constant in $\theta_\Sch$. The domain of integration in the above is the range of $\sigma_\Sch$ over which $f$ is defined, e.g.~$[0,\infty)$ for $\Phi^{(1)},\Phi^{(2)}$. The goal is to determine whether it is possible to have non-trivial initial data $\Phi^{(1)}$ on one side of the black hole while having only vanishingly small initial data $\Phi^{(2)}$ on the other side.

Before we proceed with numerical calculations, let us make some basic preliminary observations. First, note that $\Phi_{m_\Sch n_\Sch}(\sigma_\Sch)$ has a $|n_\Sch|$-order zero at $\sigma_\Sch=0$, making $\Phi_{m_\Sch n_\Sch}(\sigma_\Sch)$ linearly independent for different $|n_\Sch|$. This means that it should not be possible make $\Phi^{(2)}(t_\Sch=0)$ vanish completely while having a non-trivial $\Phi^{(1)}(t_\Sch=0)$, at least, when considering finitely many source modes. Secondly, from plotting $\Phi_{0 n_\Sch}(\sigma_\Sch)$ in Figure \ref{fig:Phi_0n_sigma}, we note that the smaller $|n_\Sch|$ functions $\Phi_{0n_\Sch}(\sigma_\Sch)$ are concentrated closer $\sigma_\Sch=0$ and differ significantly from each other; but, as $|n_\Sch|$ is increased, the $\Phi_{0n_\Sch}$ are pushed towards the boundary $\sigma_\Sch=\infty$ and become increasingly similar. A naive way to get a large $\Phi^{(1)}(t_\Sch=0)$ at the cost of a comparatively small $\Phi^{(2)}(t_\Sch=0)$ would be to just pick a large $n_\Sch$ and make $\lambda_{0,n_\Sch}\approx\lambda_{0,n_\Sch+1}\ne 0$. The $\Phi_{0,n_\Sch}(\sigma_\Sch), \Phi_{0,n_\Sch+1}(\sigma_\Sch)$ would then combine constructively in \eqref{eq:initial_data_Phi1}, but mostly cancel in \eqref{eq:initial_data_Phi2}. Thus, we should expect those sources which localize initial data to one side of the black hole to have large $|n_\Sch|$ modes and to produce initial data which are far away from the horizon.

Now, let us discuss our numerical methods for minimizing $N[\Phi^{(1)},\Phi^{(2)}]$. Let us first introduce some notation, beginning with
\begin{align*}
    I_{n_\Sch n_\Sch'}^{(\alpha)}
  \equiv& \int_0^\infty d\sigma_\Sch\; \sigma_\Sch^\alpha\sech(\sigma_\Sch)
          \left[b_{0n_\Sch} P_{-1/2}^{|n_\Sch|}(\sech(\sigma_\Sch))\right]
          \left[b_{0n_\Sch'} P_{-1/2}^{|n_\Sch'|}(\sech(\sigma_\Sch))\right], \qquad(\alpha\in\mathbb{R})
\end{align*}
which are integrals that we shall compute numerically. Note that the matrix $I^{(\alpha)}$ has all positive entries and is symmetric and positive definite since, for any complex sequence $a_{n_\Sch}$,
\begin{align*}
  \vec{a}^\dagger I^{(\alpha)} \vec{a}
  =& \sum_{n_\Sch n_\Sch'} a_{n_\Sch}^* I_{n_\Sch n_\Sch'}^{(\alpha)} a_{n_\Sch'}
  = L^2_{\sigma_\Sch}\left[
     \sqrt{\sigma_\Sch^\alpha\sech(\sigma_\Sch)} \sum_{n_\Sch} a_{n_\Sch} b_{0n_\Sch} P_{-1/2}^{|n_\Sch|}(\sech(\sigma_\Sch))
     \right],
\end{align*}
theoretically justifying taking the inverse $(I^{(\alpha)})^{-1}$. The above also gives a convenient way to write the $L^2$ norms of the fields on the two sides of the black hole:
\begin{align*}
  L^2_{\sigma_\Sch}[\Phi^{(1)}(t_\Sch=0)]
  =& \vec{\lambda}_0^\dagger I^{(0)} \vec{\lambda}_0
  \\
  L^2_{\sigma_\Sch}[\Phi^{(2)}(t_\Sch=0)]
    =& \vec{\lambda}_0^\dagger J I^{(0)} J \vec{\lambda}_0,
  & J
    \equiv& \diag(1,-1,1,-1,\ldots).
\end{align*}

We turn now to the bussiness of minimizing $N[\Phi^{(1)},\Phi^{(2)}]$. To maximize $L^2_{\sigma_\Sch}[\Phi^{(1)}(t_\Sch=0)]$ for fixed $L^2_{\sigma_\Sch}[\Phi^{(2)}(t_\Sch=0)]$, we consider the action
\begin{align*}
  S_\Lambda
  =& -L^2_{\sigma_\Sch}[\Phi^{(1)}(t_\Sch=0)]+\Lambda L^{2}_{\sigma_\Sch}[\Phi^{(2)}(t_\Sch=0)]
\end{align*}
with Lagrange multiplier $\Lambda$. Extremizing this action for some value of $\Lambda$ is a necessary condition for the maximization of $L^2_{\sigma_\Sch}[\Phi^{(1)}(t_\Sch=0)]$ with a fixed $L^2_{\sigma_\Sch}[\Phi^{(2)}(t_\Sch=0)]$. Differentiating with respect to $\vec{\lambda}_0^\dagger$, we get
\begin{align}
  0
  =& \frac{\partial S_\Lambda}{\partial \vec{\lambda}_0^\dagger}
     = [-I^{(0)}+\Lambda JI^{(0)} J] \vec{\lambda}_0, \label{eq:maximizingPhi1Phi2Ratio_generalizedEigenval}
\end{align}
a generalized eigenvalue problem\footnote{See Appendix \S\ref{sec:generalized_eigenvalue_problem}.}. This can also be turned into a standard eigenvalue problem
\begin{align}
     J (I^{(0)})^{-1} J I^{(0)} \vec{\lambda}_0
  = \Lambda \vec{\lambda}_0.
  \label{eq:maximizingPhi1Phi2Ratio_standardEigenval}
\end{align}
Note that any $\vec{\lambda}_0$ satisfying \eqref{eq:maximizingPhi1Phi2Ratio_generalizedEigenval} has
\begin{align*}
  \Lambda
  =& N[\Phi^{(1)},\Phi^{(2)}],
\end{align*}
so the maximized ratio  $N[\Phi^{(1)},\Phi^{(2)}]$ is just the maximum eigenvalue of \eqref{eq:maximizingPhi1Phi2Ratio_generalizedEigenval}.

Due to \eqref{eq:Phi_mn_sign_of_n},
\begin{align*}
  I^{(\alpha)}_{n_\Sch,n_\Sch'}
  =&I^{(\alpha)}_{|n_\Sch|,|n_\Sch'|},
\end{align*}
so it suffices to consider only the part of the matrices with $n_\Sch,n_\Sch'\ge 0$, e.g.
\begin{align*}
  \sum_{n_\Sch'\in\mathbb{Z}}  I^{(\alpha)}_{n_\Sch n_\Sch'} \lambda_{0n_\Sch'}
  =& \sum_{n_\Sch'\ge 0}  I^{(\alpha)}_{n_\Sch n_\Sch'} \begin{cases}
    \lambda_{00} &\text{if $n_\Sch'=0$} \\
    \lambda_{0,n_\Sch'}+\lambda_{0,-n_\Sch'} &\text{otherwise}
  \end{cases}.
\end{align*}
In practice, we will bound $|n_\Sch|,|n_\Sch'|$ from above:
\begin{align*}
  |n_\Sch|,|n_\Sch'| \le& n_{\max}.
\end{align*}
It will be interesting to see whether the maximized ratio $N[\Phi^{(1)},\Phi^{(2)}]$ increases without bound as $n_{\max}$ is increased.

So far, we have neglected the constraints \eqref{eq:lambdaConstraint_timeReversal}-\eqref{eq:lambdaConstraint_real} on the sources. It is trivial to impose \eqref{eq:lambdaConstraint_timeReversal} because the generalized eigenvalue problem \eqref{eq:maximizingPhi1Phi2Ratio_generalizedEigenval} involves real symmetric matrices so  $\vec{\lambda}_0$ can be chosen to be real. To impose \eqref{eq:lambdaConstraint_vanishAtInitialTime}, it suffices to modify $I^{(0)}$ by subtracting the $0$th and $1$st rows and columns from all other even and odd rows and columns respectively, then considering only the vector space occupied by $\lambda_{0,n_\Sch}+\lambda_{0,-n_\Sch}$ for $n\ge 2$. Then, $\lambda_{00},\lambda_{0,1}+\lambda_{0,-1}$ are determined by
\begin{align*}
  \lambda_{00} =& -\sum_{n_\Sch=2,4,\ldots}(\lambda_{0,n_\Sch}+\lambda_{0,-n_\Sch}),
  & \lambda_{0,1}+\lambda_{0,-1}=&  -\sum_{n_\Sch=3,5,\ldots}(\lambda_{0,n_\Sch}+\lambda_{0,-n_\Sch}).
\end{align*}
Finally, the constraint \eqref{eq:lambdaConstraint_real}, coming from assuming $\lambda(\tau_\Sch)\in\mathbb{R}$, simply kills the degeneracy left from the fact that the variational problem only determines the combinations $\lambda_{00},\lambda_{0,n_\Sch}+\lambda_{0,-n_\Sch}$.

Before presenting our results, let us briefly remark upon some logistics of the numerical computation. Since solving either the generalized eigenvalue problem \eqref{eq:maximizingPhi1Phi2Ratio_generalizedEigenval} or the standard eigenvalue problem \eqref{eq:maximizingPhi1Phi2Ratio_standardEigenval} requires inverting $I^{(0)}$, we must ensure that $I^{(0)}$ is calculated with enough numerical precision that the smallest eigenvalue of $I^{(0)}$ can be reliably found. As $n_{\max}$ is increased, $I^{(0)}$ becomes exponentially close to being singular, with its smallest eigenvalue of approximate order $10^{-3n_{\max}/2}$. On the other hand, the entries of $I^{(0)}$ remain relatively large, e.g.~$I^{(0)}_{100,100}\approx 0.005$. Then, to get $j$ digits of precision in the smallest eigenvalue of $I^{(0)}$, we need to calculate the entries of $I^{(0)}$ to about $(j+3n_{\max}/2)$-many digits of precision. This high-precision computation of numerical integrals quickly becomes a bottleneck for large $n_{\max}$ calculations. We will work with $n_{\max}$ up to a maximum of $100$. To be safe, we have calculated the numerical integrals $I^{(0)}_{n_\Sch n_\Sch'}$ to $175$ digits of precision.

\begin{figure}
  \centering
  \begin{subfigure}[c]{0.475\textwidth}
    \includegraphics[width=\textwidth]{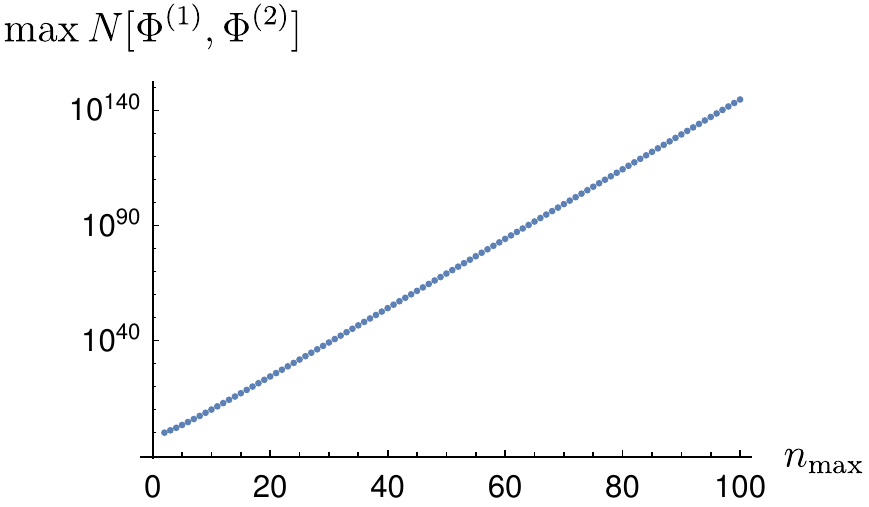}
    \caption{Maximized ratio of the $L^2_{\sigma_\Sch}$ norms of initial data on the two sides of the black hole.}
    \label{fig:Phi1Phi2Ratio_maximizingPhi1Phi2Ratio}
  \end{subfigure}
  \hfill
  \begin{subfigure}[c]{0.475\textwidth}
    \includegraphics[width=\textwidth]{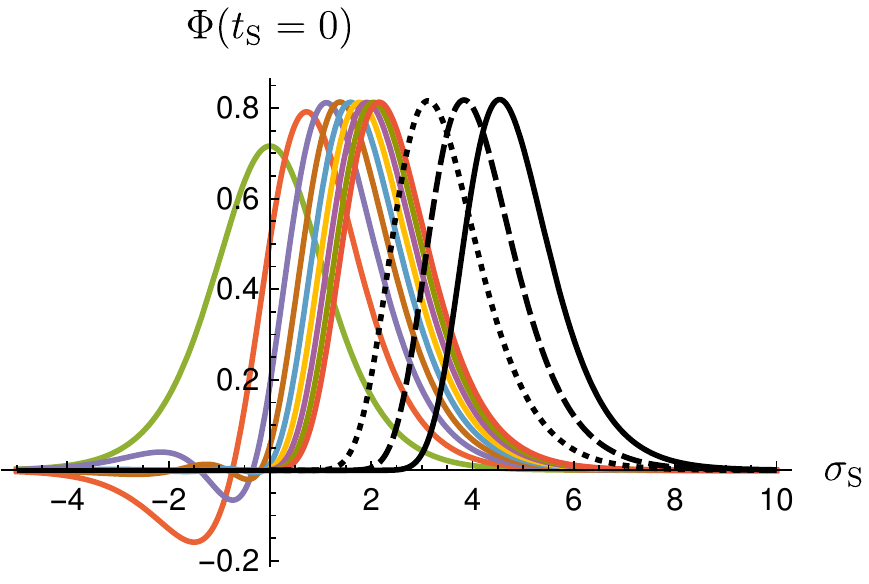}
    \caption{Initial data $\Phi^{(1)}(t_\Sch=0,\sigma_\Sch)$  (plotted for $\sigma_\Sch\ge 0$) and $\Phi^{(2)}(t_\Sch=0,-\sigma_\Sch)$ (plotted for $\sigma_\Sch\le 0$) on the two sides of the black hole. The coloured curves correspond to $n_{\max}=2,3,\ldots,10$ and the black curves correspond to $n_{\max}=25,50,100$.}
    \label{fig:Phi1_Phi2_maximizingPhi1Phi2Ratio}
  \end{subfigure}
  \par\bigskip
  \begin{subfigure}[c]{0.475\textwidth}
    \includegraphics[width=\textwidth]{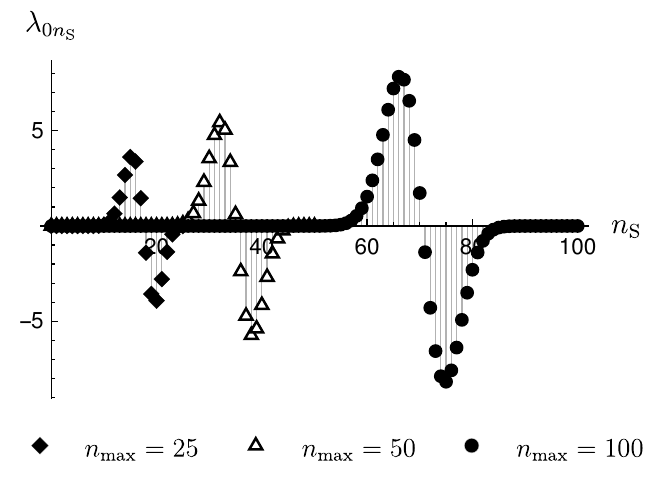}
    \caption{Fourier coefficients $\lambda_{0n_\Sch}$ of sources. From assuming $\lambda(\tau_\Sch)\in\mathbb{R}$, we have $\lambda_{0,-n_\Sch}=\lambda_{0,n_\Sch}$.}
    \label{fig:lambda0n_maximizingPhi1Phi2Ratio}
  \end{subfigure}
  \hfill
  \begin{subfigure}[c]{0.475\textwidth}
    \includegraphics[width=\textwidth]{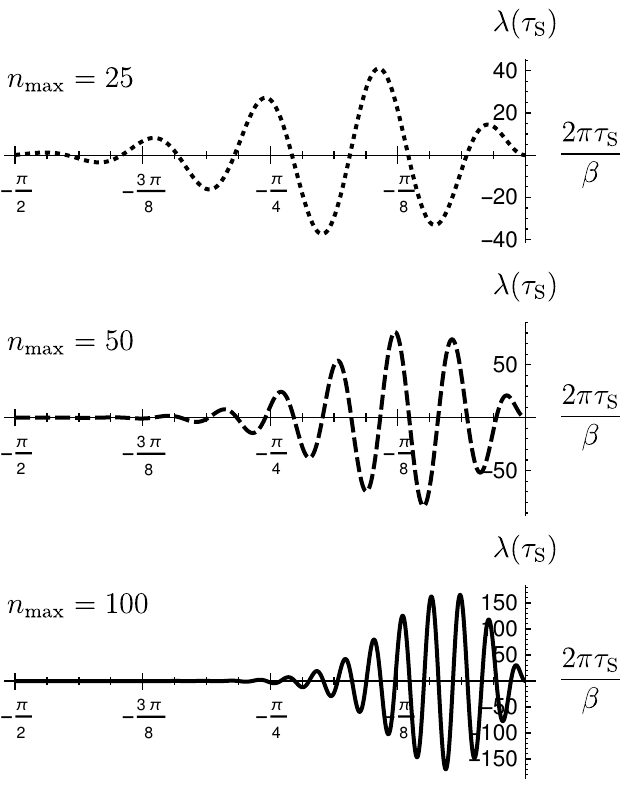}
    \caption{Sources $\lambda$ as functions of Euclidean time $\tau_\Sch$. There is a $\beta$-periodicity in $\tau_\Sch$ and $\lambda(\tau_\Sch)=\lambda(-\tau_\Sch)$.}
    \label{fig:lambda_maximizingPhi1Phi2Ratio}
  \end{subfigure}
  \caption{Maximization of the ratio $N[\Phi^{(1)},\Phi^{(2)}]$ of $L^2_{\sigma_\Sch}$ norms of the initial data $\Phi^{(1)}(t_\Sch=0),\Phi^{(2)}(t_\Sch=0)$ on the two sides of the black hole. The sources are assumed to be real functions of Euclidean time.}
  \label{fig:maximizingPhi1Phi2Ratio}
\end{figure}

Our results are shown in Figure \ref{fig:maximizingPhi1Phi2Ratio}. We see from Figure \ref{fig:Phi1Phi2Ratio_maximizingPhi1Phi2Ratio} that the $L^2_{\sigma_\Sch}$ ratio of the initial data on the two sides of the black hole can be made exponentially large as the cutoff $n_{\max}$ is increased. Moreover, Figure \ref{fig:Phi1_Phi2_maximizingPhi1Phi2Ratio} vindicates our earlier guess that the initial data which maximize the $L^2_{\sigma_\Sch}$ ratio
become shifted further away from the horizon as $n_{\max}$ is increased. Additionally, Figure \ref{fig:lambda0n_maximizingPhi1Phi2Ratio} tells us that the sources responsible for those initial data are indeed composed mostly of large $n_\Sch$ Fourier modes, though it is interesting to see that the peaks of $\lambda_{0n_\Sch}$ are several half-widths away from the largest allowed frequency $n_{\max}$. As a function of Euclidean time, $\lambda$ in Figure \ref{fig:lambda_maximizingPhi1Phi2Ratio} resembles a wave-packet that gets squeezed towards $\tau_\Sch=0$ as $n_{\max}$ is increased.

\subsection{Localizing initial data in $\sigma_\Sch$}
While we have seen that it is possible to produce perturbations that are well-localized in one asymptotic region of the two-sided black hole, we may further wish to localize perturbations at some particular radius. To understand to what extent this is possible, we consider the minimization of the variance
\begin{align*}
  V[\Phi](\sigma_0)
  \equiv& \frac{\int_{-\infty}^\infty d\sigma_\Sch\; (\sigma_\Sch-\sigma_0)^2 [\Phi(t_\Sch=0)]^2}
          {L_{\sigma_\Sch}^2[\Phi(t_\Sch=0)]}
\end{align*}
of the initial data
\begin{align*}
  \Phi(t_\Sch=0,\sigma_\Sch) \equiv& \begin{cases}
    \Phi^{(1)}(t_\Sch=0,\sigma_\Sch) &\text{if $\sigma_\Sch\ge 0$} \\
    \Phi^{(2)}(t_\Sch=0,-\sigma_\Sch) &\text{if $\sigma_\Sch<0$}
  \end{cases}.
\end{align*}
This definition is reasonable because $\Phi^{(1)}(t_\Sch=0,\sigma_\Sch)$ always connects smoothly with $\Phi^{(2)}(t_\Sch=0,-\sigma_\Sch)$ at the horizon $\sigma_\Sch=0$, at least for a finite UV cutoff $n_{\max}$ on the sources.

\begin{figure}
  \centering
  \begin{subfigure}[c]{\textwidth}
    \includegraphics[width=\textwidth]{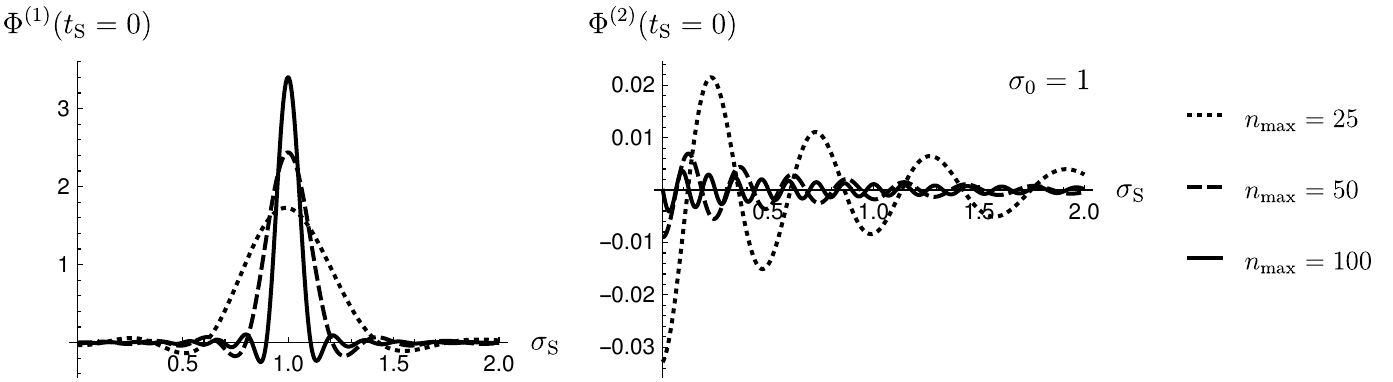}
    \caption{Localized initial data which minimize the variance $V[\Phi]$ about $\sigma_0=1$, plotted for UV cutoffs $n_{\max}=25,50,100$ on the sources.}
    \label{fig:Phi1_Phi2_localizingPhi}
  \end{subfigure}
  \par\bigskip
  \begin{subfigure}[c]{0.475\textwidth}
    \includegraphics[width=\textwidth]{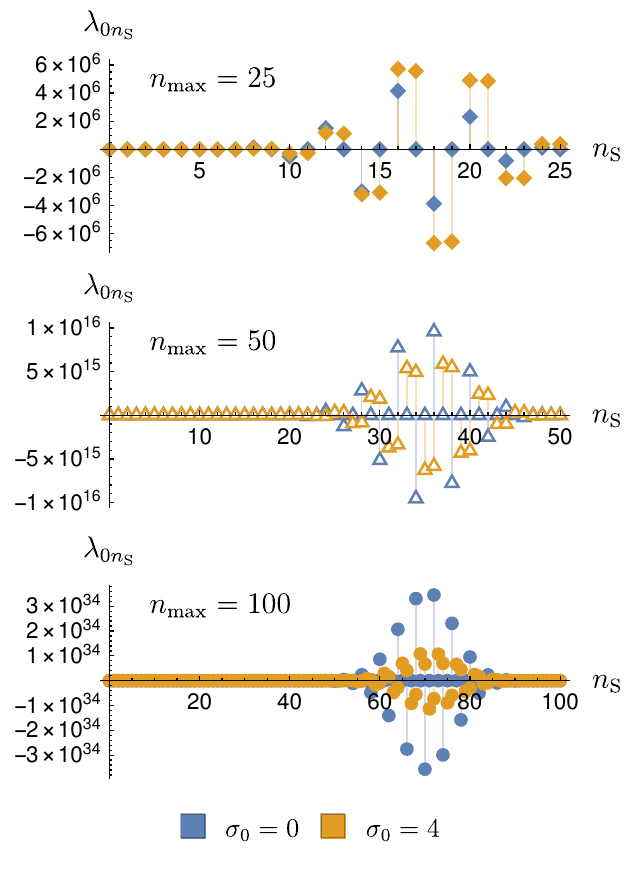}
    \caption{Fourier coefficients $\lambda_{0n_\Sch}$ of sources which localize initial data on the horizon $\sigma_0=0$ and on one side of the black hole $\sigma_0=4$.}
    \label{fig:lambda0n_localizingPhi}
  \end{subfigure}
  \hfill
  \begin{subfigure}[c]{0.475\textwidth}
    \includegraphics[width=\textwidth]{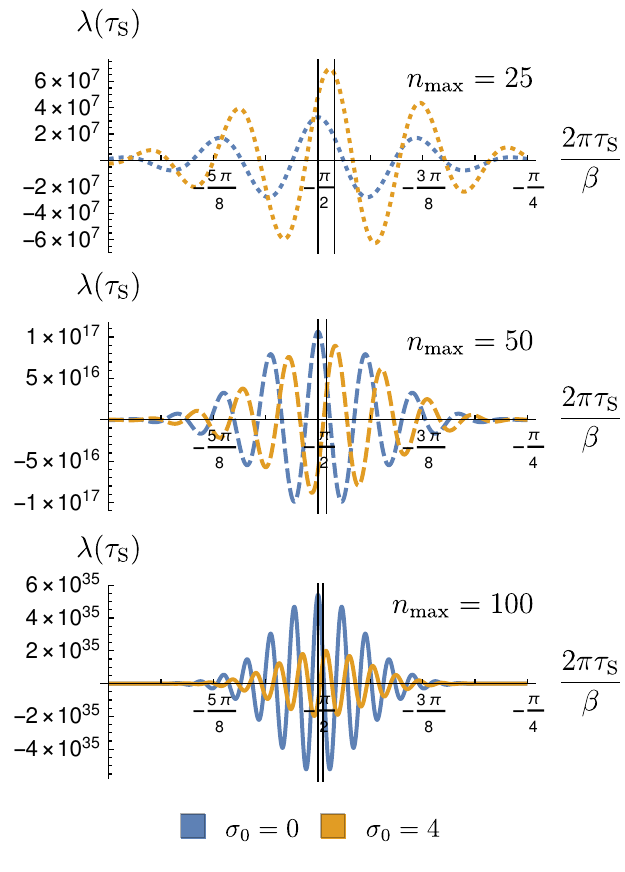}
    \caption{Sources $\lambda$ of Figure \ref{fig:lambda0n_localizingPhi} plotted as functions of Euclidean time $\tau_\Sch$. The black vertical lines mark the $\tau_\Sch$ values which maximize the amplitude of the oscillations in $\lambda$, with the left and right lines corresponding to $\sigma_0=0,4$ respectively.}
    \label{fig:lambda_localizingPhi}
  \end{subfigure}
  \caption{Initial data, localized by the minimization of variance $V[\Phi]$ about $\sigma_0$, and sources that produce such initial data.}
  \label{fig:sourcesAndInitialData_localizingPhi}
\end{figure}

\begin{figure}
  \centering
  \begin{subfigure}[c]{0.475\textwidth}
    \includegraphics[width=\textwidth]{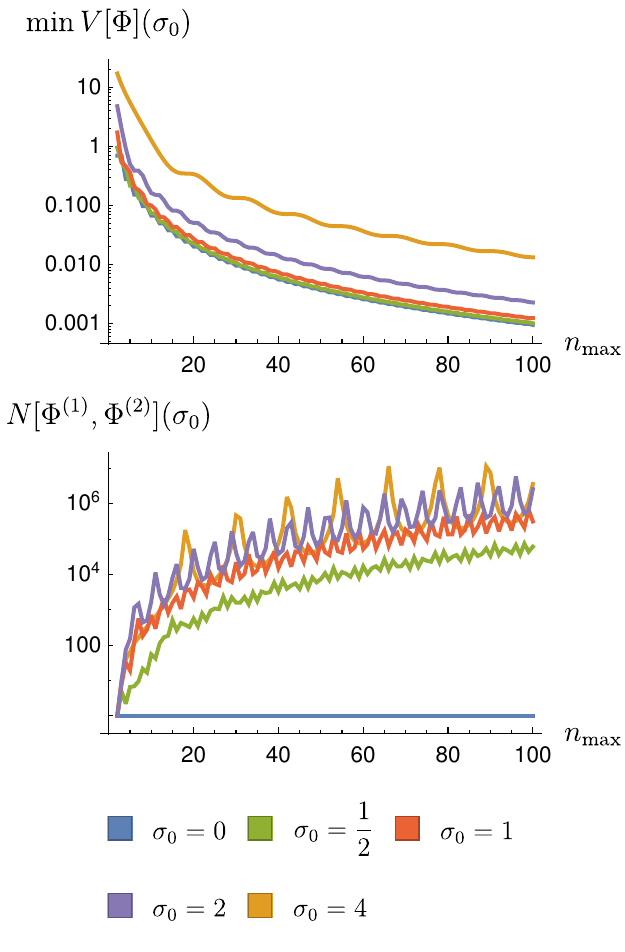}
    \caption{Minimized variance $V[\Phi]$ of the initial data and the ratio $N[\Phi^{(1)},\Phi^{(2)}]$ of $L^2_{\sigma_\Sch}$ norms between the two sides of the black hole, plotted against the UV cutoff $n_{\max}$ on the source modes.}
    \label{fig:Xi_Phi1Phi2Ratio_localizingPhi}
  \end{subfigure}
  \hfill
  \begin{subfigure}[c]{0.475\textwidth}
    \includegraphics[width=\textwidth]{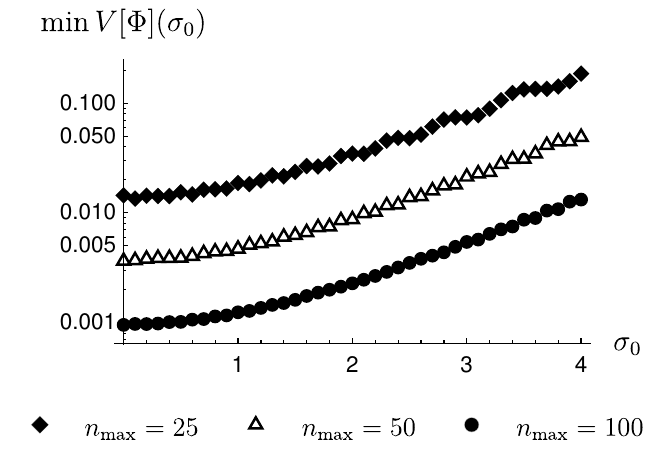}
    \caption{Minimized variance $V[\Phi]$ plotted against the position $\sigma_0$ about which the initial data is localized.}
    \label{fig:XiVersussigma0_localizingPhi}
  \end{subfigure}
  \caption{Dependence of the localization of initial data $\Phi$ on the UV cutoff $n_{\max}$ of the sources and the position $\sigma_0$ of localization.}
  \label{fig:varianceAndL2Ratio_localizingPhi}
\end{figure}

A necessary condition for the minimization of $V[\Phi](\sigma_0)$ is that the action
\begin{align*}
  S_\Xi
  =& \int_{-\infty}^\infty d\sigma_\Sch\; (\sigma_\Sch-\sigma_0)^2 [\Phi(t_\Sch=0)]^2
     -\Xi L_{\sigma_\Sch}^2[\Phi(t_\Sch=0)] \\
  =& \vec{\lambda}_0^\dagger (M^{(2)}-\Xi M^{(0)})\vec{\lambda}_0 \\
  M^{(2)}
  \equiv& (I^{(2)}-2\sigma_0 I^{(1)} + \sigma_0^2 I^{(0)})
          + J(I^{(2)}+2\sigma_0 I^{(1)} + \sigma_0^2 I^{(0)})J \\
  M^{(0)}
  \equiv& I^{(0)}+JI^{(0)}J.
\end{align*}
must be extremized for some value of the Lagrange multiplier $\Xi$:
\begin{align}
  0=& \frac{\partial S_{\Xi\xi}}{\partial \vec{\lambda}_0^\dagger}
      = (M^{(2)}-\Xi M^{(0)}) \vec{\lambda}_0,
  &
  &\text{i.e.,}
    &
    (M^{(0)})^{-1} M^{(2)} \vec{\lambda}_0
      =& \Xi \vec{\lambda}_0 .
      \label{eq:localizingPhi}
\end{align}
Note that solutions to \eqref{eq:localizingPhi} satisfy
\begin{align*}
  V[\Phi](\sigma_0) =&  \Xi.
\end{align*}
To minimize $V[\Phi](\sigma_0)$ over all possible sources, it suffices to find the minimum eigenvalue of \eqref{eq:localizingPhi}. As decribed previously in \S\ref{sec:maximizingPhi1Phi2Ratio}, we can restrict ourselves to considering only the $n_\Sch,n_\Sch'\ge 0$ entries of matrices. Using the same methods as in \S\ref{sec:maximizingPhi1Phi2Ratio}, we again impose the constraints \eqref{eq:lambdaConstraint_timeReversal}-\eqref{eq:lambdaConstraint_real} on the sources. The results are plotted in Figures \ref{fig:sourcesAndInitialData_localizingPhi} and \ref{fig:varianceAndL2Ratio_localizingPhi}.

Figure \ref{fig:sourcesAndInitialData_localizingPhi} shows examples of initial data localized through the minimization of $V[\Phi](\sigma_0)$ and sources which produce such initial data. In Figure \ref{fig:Phi1_Phi2_localizingPhi}, we see that $\Phi$ can be made increasingly localized as the UV cutoff $n_{\max}$ on the sources is raised. Not surprisingly, we also see that the minimization of $V[\Phi](\sigma_0)$ about a $\sigma_0>0$ on one side of the black hole automatically reduces the size of the initial data on the other side of the black hole. Plotting the Fourier coefficients of the sources in Figure \ref{fig:lambda0n_localizingPhi}, we find that the envelope of $|\lambda_{0n}|$ appears to be single-peaked distribution, centered similarly to the Fourier coefficients in Figure \ref{fig:lambda0n_maximizingPhi1Phi2Ratio}. In Figure \ref{fig:lambda_localizingPhi}, we find that $\lambda(\tau_\Sch)$ resembles a wavepacket that becomes increasingly localized as $n_{\max}$ is raised. In those plots, we have added a vertical line to mark the maximum of each wavepacket's amplitude (determined by considering $\lambda(\tau_\Sch)-\lambda_{00}$ and its Hilbert transform). Somewhat surprisingly, this maximum moves towards $\tau=-\beta/4$ as $n_{\max}$ is raised, even when localizing $\Phi$ about $\sigma_0>0$.

Thus, the sources required to produce very localized perturbations on one side of the black hole are not localized to the corresponding side of the path integral cylinder, but rather concentrated around the middle point $\tau=-\beta/4$.

We further explore the dependence on the sources' UV cutoff $n_{\max}$ and the location $\sigma_0$ of localization in Figure \ref{fig:varianceAndL2Ratio_localizingPhi}. In Figure \ref{fig:Xi_Phi1Phi2Ratio_localizingPhi}, we plot the minimized variance $V[\Phi](\sigma_0)$ and corresponding ratio $N[\Phi^{(1)},\Phi^{(2)}]$ of $L_{\sigma_\Sch}^2$ norms against $n_{\max}$. Consistent with our discussion of Figure \ref{fig:Phi1_Phi2_localizingPhi}, it indeed appears that $V[\Phi](\sigma_0)$ can be made arbitrarily small as $n_{\max}$ is raised and, for $\sigma_0>0$, this results in an increasingly large ratio $N[\Phi^{(1)},\Phi^{(2)}]$ of $L_{\sigma_\Sch}^2$ norms between the two sides of the black hole. Moreover, for a fixed $n_{\max}$, we find that it becomes more difficult to localize the initial data $\Phi$ as we move away from the horizon $\sigma_0=0$. This is verified in Figure \ref{fig:XiVersussigma0_localizingPhi}, where we plot the minimized variance $V[\Phi](\sigma_0)$ against $\sigma_0$.

\subsubsection{Fixing ratio of $L^2$ norms of initial data and sources}
\label{sec:localizingPhiFixingPhilambdaRatio}
In this section, we wish to minimize $V[\Phi](\sigma_0)$ for a fixed ratio between the $L^2$ norms of the sources and initial data
\begin{align*}
  N[\Phi,\lambda]\equiv& \frac{L^2_{\sigma_\Sch}[\Phi(t_\Sch=0)]}{L^2_{\tau_\Sch}[\lambda]} \\
  L^2_{\tau_\Sch}[\lambda] \equiv& \frac{2}{\beta} \int_0^{\beta/2} d\tau_\Sch\; |\lambda(\tau_\Sch)|^2
                                   = \sum_n |\lambda_{n_\Sch}|^2.
\end{align*}
Specifically, we would like to see whether the localization of $\Phi$ requires increasingly large sources, as observed in the pure $\AdS$ case studied by \cite{Marolf:2017kvq}.

A necessary condition for the minimization of $V[\Phi](\sigma_0)$ is that the action
\begin{align*}
  S_{\Xi\xi}
  =& \int_{-\infty}^\infty d\sigma_\Sch\; (\sigma_\Sch-\sigma_0)^2 [\Phi(t_\Sch=0)]^2
     -\Xi L_{\sigma_\Sch}^2[\Phi(t_\Sch=0)]
     +\xi L_{\tau_\Sch}^2[\lambda] \\
  =& \vec{\lambda}_0^\dagger (M^{(2)}-\Xi M^{(0)} + \xi \mathds{1})\vec{\lambda}_0.
\end{align*}
must be extremized for some values of the Lagrange multipliers $\Xi,\xi$:
\begin{align}
  0=& \frac{\partial S_{\Xi\xi}}{\partial \vec{\lambda}_0^\dagger}
      = (M^{(2)}-\Xi M^{(0)}+\xi \mathds{1}) \vec{\lambda}_0.
      \label{eq:localizingPhiFixingPhilambdaRatio}
\end{align}
For a given $\Xi$, this is an eigenvalue problem. Note that solutions to \eqref{eq:localizingPhiFixingPhilambdaRatio} satisfy
\begin{align}
  V[\Phi](\sigma_0) =&  \Xi-\frac{\xi}{N[\Phi,\lambda]}.
                       \label{eq:VPhi_localizingPhiFixingPhilambdaRatio}
\end{align}
Our strategy for minimizing $V[\Phi](\sigma_0)$ for fixed $N[\Phi,\lambda]$ will be as follows. First, we will solve the eigenvalue problem \eqref{eq:localizingPhiFixingPhilambdaRatio} for a large set of $\Xi$-values. Each eigenvector will contribute a point on the $N[\Phi,\lambda], V[\Phi](\sigma_0)$ plane. The curve plotting the minimized $V[\Phi](\sigma_0)$ for fixed $N[\Phi,\lambda]$ will then be taken to be the lower boundary of that set of points.

\begin{figure}
  \centering
  \begin{subfigure}[c]{0.475\textwidth}
    \includegraphics[width=\textwidth]{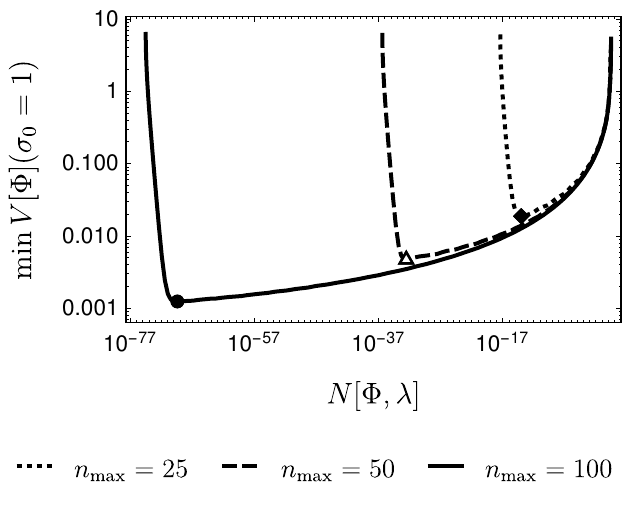}
    \caption{Varying the UV cutoff $n_{\max}$ on the source modes.}
    \label{fig:Xi_Phi1Phi2Ratio_localizingPhiFixingPhilambdaRatio}
  \end{subfigure}
  \hfill
  \begin{subfigure}[c]{0.475\textwidth}
    \includegraphics[width=\textwidth]{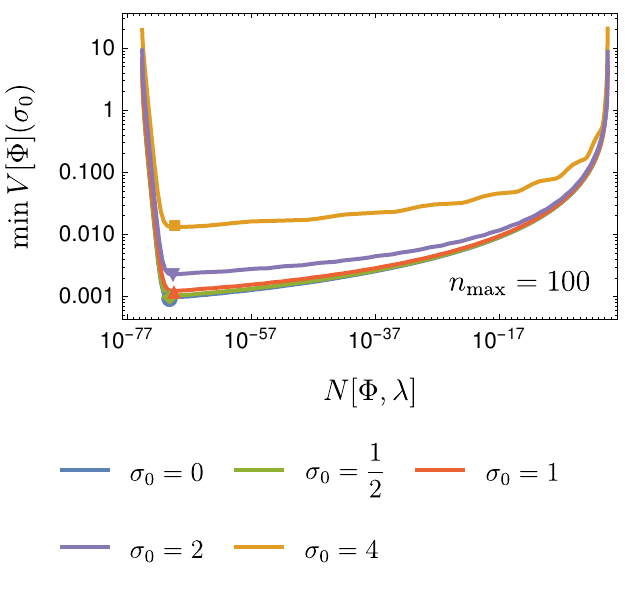}
    \caption{Varying the location $\sigma_0$ about which $\Phi$ is localized.}
    \label{fig:XiVersussigma0_localizingPhiFixingPhilambdaRatio}
  \end{subfigure}
  \caption{Minimization of the variance $V[\Phi]$ of the initial data for fixed ratios of $L^2$ norms between the initial data $\Phi$ and the sources $\lambda$. Shapes mark the absolute minima of $V[\Phi]$.}
  \label{fig:varianceAndL2Ratio_localizingPhiFixingPhilambdaRatio}
\end{figure}

The results are plotted in Figure \ref{fig:varianceAndL2Ratio_localizingPhiFixingPhilambdaRatio}. In Figure \ref{fig:Xi_Phi1Phi2Ratio_localizingPhiFixingPhilambdaRatio}, we plot the minimized variance $V[\Phi]$ against $N[\Phi,\lambda]$ for $\sigma_0=1$ and $n_{\max}=25,50,100$. For each $n_{\max}$, we see that the curve consists of two parts, divided by the minimum of the curve. The part to left of the minimum, gets pushed towards ever smaller $N[\Phi,\lambda]$ as the UV cutoff $n_{\max}$ on the sources is raised, suggesting that in the $n_{\max}\to\infty$ limit, this part of the curve disappears. In contrast, the part to the right of the minimum converges to a fixed increasing curve as $n_{\max}$ is raised. From this, we conclude that in the $n_{\max}\to\infty$ limit, the minimized variance $V[\Phi]$ converges to an increasing function of $N[\Phi,\lambda]$. Thus, as in the pure $\AdS$ case studied in \cite{Marolf:2017kvq}, the localization of initial data for the BTZ black hole requires increasingly large sources. In Figure \ref{fig:XiVersussigma0_localizingPhiFixingPhilambdaRatio}, we superpose the minimized $V[\Phi](\sigma_0)$ versus $N[\Phi,\lambda]$ plots for several $\sigma_0$, finding, as suggested earlier, that it becomes more difficult to localize initial data as one moves away from the horizon.

\subsubsection{Fixing ratio of $L_{\sigma_\Sch}^2$ norms of $\Phi^{(1)}(t_\Sch=0),\Phi^{(2)}(t_\Sch=0)$}
In this section, we wish to minimize the variance
\begin{align*}
  V[\Phi^{(1)}](\sigma_0)
  \equiv& \frac{\int_0^\infty d\sigma_\Sch\; (\sigma_\Sch-\sigma_0)^2 [\Phi^{(1)}(t_\Sch=0)]^2}
          {L_{\sigma_\Sch}^2[\Phi^{(1)}(t_\Sch=0)]}
\end{align*}
of the initial data $\Phi^{(1)}(t_\Sch=0)$ on one side of the black hole for a fixed $N[\Phi^{(1)},\Phi^{(2)}]$. A necessary condition is that the action
\begin{align*}
  S_{\Xi\Lambda}
  =& \int_0^\infty d\sigma_\Sch\; (\sigma_\Sch-\sigma_0)^2 [\Phi^{(1)}(t_\Sch=0)]^2
     -\Xi L_{\sigma_\Sch}^2[\Phi^{(1)}(t_\Sch=0)]
     +\Lambda L_{\sigma_\Sch}^2[\Phi^{(2)}(t_\Sch=0)] \\
  =& \vec{\lambda}_0^\dagger [I^{(2)}-2\sigma_0 I^{(1)} + \sigma_0^2 I^{(0)}] \vec{\lambda}_0
     -\Xi \vec{\lambda}_0^\dagger I^{(0)}\vec{\lambda}_0
     +\Lambda \vec{\lambda}_0^\dagger JI^{(0)}J \vec{\lambda}_0.
\end{align*}
must be extremized for some values of the Lagrange multipliers $\Xi,\Lambda$:
\begin{align}
  0=& \frac{\partial S_{\Xi\Lambda}}{\partial \vec{\lambda}_0^\dagger}
      = [I^{(2)}-2\sigma_0 I^{(1)} + (\sigma_0^2-\Xi) I^{(0)} + \Lambda J I^{(0)} J] \vec{\lambda}_0.
      \label{eq:localizingPhi1FixingPhi1Phi2Ratio_generalizedEigenval}
\end{align}
For a given $\Xi$, this is a generalized eigenvalue problem. Alternatively, the above can be recast into a standard eigenvalue problem:
\begin{align}
  J (I^{(0)})^{-1} J [-I^{(2)}+2\sigma_0 I^{(1)} + (\Xi-\sigma_0^2) I^{(0)}]\vec{\lambda}_0
  =& \Lambda  \vec{\lambda}_0.
     \label{eq:localizingPhi1FixingPhi1Phi2Ratio_standardEigenval}
\end{align}
Note that solutions to \eqref{eq:localizingPhi1FixingPhi1Phi2Ratio_generalizedEigenval} satisfy
\begin{align}
  V[\Phi^{(1)}](\sigma_0) =&  \Xi - \frac{\Lambda}{N[\Phi^{(1)},\Phi^{(2)}]}.
                            \label{eq:VPhi1_localizingPhi1FixingPhi1Phi2Ratio}
\end{align}
Our strategy for determining the minimized $V[\Phi^{(1)}]$ versus $N[\Phi^{(1)},\Phi^{(2)}]$ curve will be completely analogous to \S\ref{sec:localizingPhiFixingPhilambdaRatio}, with \eqref{eq:localizingPhi1FixingPhi1Phi2Ratio_generalizedEigenval} and \eqref{eq:VPhi1_localizingPhi1FixingPhi1Phi2Ratio} replacing \eqref{eq:localizingPhiFixingPhilambdaRatio} and \eqref{eq:VPhi_localizingPhiFixingPhilambdaRatio}.

\begin{figure}
  \centering
  \begin{subfigure}[c]{0.475\textwidth}
    \includegraphics[width=\textwidth]{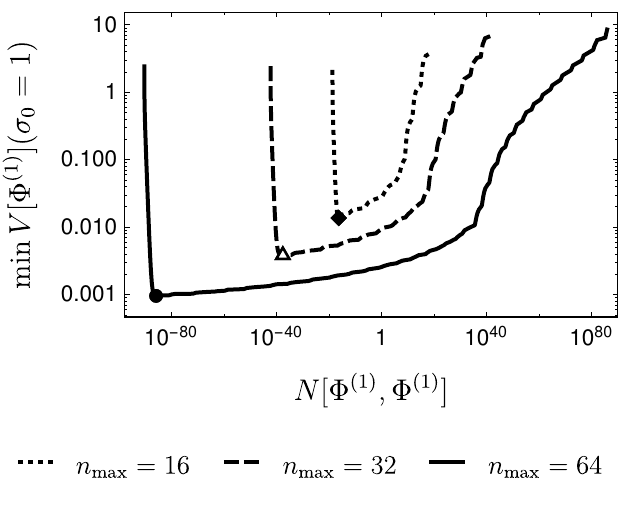}
    \caption{Varying the UV cutoff $n_{\max}$ on the source modes.}
    \label{fig:Xi_Phi1Phi2Ratio_localizingPhi1FixingPhi1Phi2Ratio}
  \end{subfigure}
  \hfill
  \begin{subfigure}[c]{0.475\textwidth}
    \includegraphics[width=\textwidth]{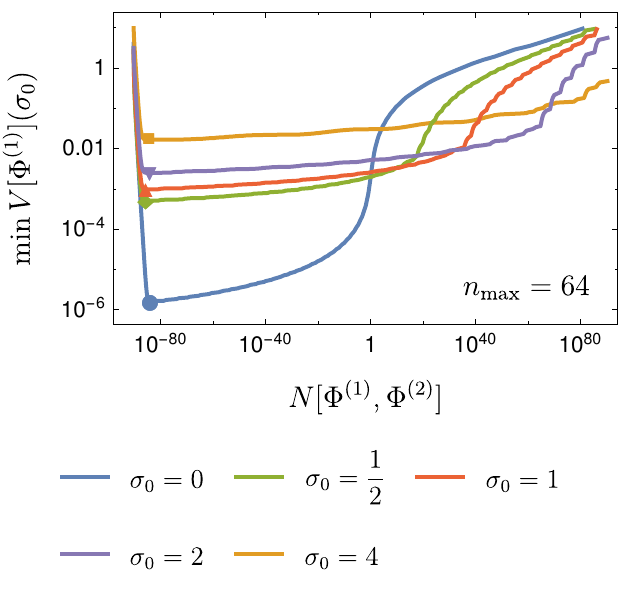}
    \caption{Varying the location $\sigma_0$ about which $\Phi^{(1)}$ is localized.}
    \label{fig:XiVersussigma0_localizingPhi1FixingPhi1Phi2Ratio}
  \end{subfigure}
  \caption{Minimization of the variance $V[\Phi^{(1)}]$ of the initial data on one side of the black hole for fixed ratios of $L^2_{\sigma_\Sch}$ norms between the initial data $\Phi^{(1)},\Phi^{(2)}$ on the two sides of the black hole. Shapes mark the absolute minima of $V[\Phi^{(1)}]$.}
  \label{fig:varianceAndL2Ratio_localizingPhi1FixingPhi1Phi2Ratio}
\end{figure}

Our results\footnote{The calculations involved in the minimization of $V[\Phi^{(1)}]$ for fixed $N[\Phi^{(1)},\Phi^{(2)}]$ seem to require even greater precision in the $I^{(\alpha)}$ matrices. To get the results in Figure \ref{fig:varianceAndL2Ratio_localizingPhi1FixingPhi1Phi2Ratio}, we calculated $I^{(\alpha)}$ to $200$ digits of precision.}  are shown in Figure \ref{fig:varianceAndL2Ratio_localizingPhi1FixingPhi1Phi2Ratio}. In Figure \ref{fig:Xi_Phi1Phi2Ratio_localizingPhi1FixingPhi1Phi2Ratio}, we plot the minimized variance $V[\Phi^{(1)}]$ against the ratio $N[\Phi^{(1)},\Phi^{(2)}]$ of $L_{\sigma_\Sch}^2$ norms for $\sigma_0=1$ and $n_{\max}=16,32,64$. As $n_{\max}\to\infty$, it appears that the allowed region in the $N[\Phi^{(1)},\Phi^{(2)}], V[\Phi^{(1)}]$ plane (i.e.~the region above each curve in Figure \ref{fig:Xi_Phi1Phi2Ratio_localizingPhi1FixingPhi1Phi2Ratio}) expands to cover the entire positive qudrant of the plane. This suggests that when the UV cutoff on the sources is removed, it is possible to produce arbitrarily localized initial data $\Phi^{(1)}$ on one side of the black hole while having initial data $\Phi^{(2)}$ of vanishingly small relative size on the other. In Figure \ref{fig:XiVersussigma0_localizingPhi1FixingPhi1Phi2Ratio}, we superpose the minimized $V[\Phi^{(1)}]$ versus $N[\Phi^{(1)},\Phi^{(2)}]$ plots for $\sigma_0=0,1/2,1,2,4$. The curves for $\sigma_0=1/2,1,2$ have the same overall features: to the left of the absolute minimum of $V[\Phi^{(1)}]$, there is a nearly vertical section indicating an approach towards a minimum $N[\Phi^{(1)},\Phi^{(2)}]$ value; to the right, there is first a section where the minimized variance $V[\Phi^{(1)}]$ increases slowly as a function of $N[\Phi^{(1)},\Phi^{(2)}]$, then an abrupt transition to a more rapid increase. Naturally, the curve corresponding to localization on the horizon $\sigma_0=0$ differs significantly from the others. The peculiarity of the $\sigma_0=4$ curve can probably be blamed on the fact that the cutoff $n_{\max}=64$ is too low to adequately localize $\Phi^{(1)}$ so far away from the horizon --- in Figure \ref{fig:Phi_0n_sigma}, note that $\Phi_{0,n\ge 64}$ should make appreciable contributions at $\sigma=4$.

\section{Summary and outlook}
In this note, we have established a holographic map between sources perturbing the path integral construction the CFT thermofield double state and the initial data of a scalar field living on an extended BTZ black hole geometry. In Euclidean signature, the map identifies Fourier modes of the sources and Schwarzschild modes of the bulk scalar. Additionally, we have numerically probed the extent to which initial data can be localized to one exterior and to a fixed spatial radius. Our conclusion is that, without other constraints, this can be done arbitrarily well. When the goal is purely to maximize the size of initial data on one exterior relative to the other, the optimal sources were naturally found to be concentrated towards the corresponding edge of the path integral half-cylinder. Surprisingly however, to localize initial data to a given radius on one exterior, the optimal choice of sources is in fact concentrated towards the middle of the path integral half-cylinder. A general feature we have encountered is that localization of initial data, either to a black hole exterior or to a fixed radius, involves bulk scalar modes of arbitrarily high frequency as it is only these modes which have concentrated support at large radius. Correspondingly, the requisite sources involve high frequency Fourier modes in Euclidean time. Moreover, we have found that greater localization of initial data generally entails a larger ratio between the sizes of the sources and initial data. As in the pure AdS case \cite{Marolf:2004fy}, this implies that for sources to remain perturbatively small, localization can only be achieved for small initial data.

In future work, it would be interesting to understand better the behaviour of the Lorentzian solutions, in particular to look at how the perturbations on the initaial data slice perturb the interior spacetime behind the black hole horizon. It would also be interesting to understand whether the qualitative lessons we have learned in the 2+1 dimensional case extend to higher dimensions, but this may be significantly more difficult since we don't expect an analytic solution for the modes perturbing higher-dimensional Schwarzschild spacetimes.

\acknowledgments
We thank Don Marolf for discussions. This work was supported by a Science Undergraduate Research Experience (SURE) award from the University of British Columbia and by the Simons Foundation.

\appendix
\section{Pure $\AdS$ metrics}
In this section, we collect the various  metrics of pure $\AdS$ and state the relationships between the various coordinate systems (see \cite{Aharony:1999ti}). We can think of $\AdS_{d+1}$ as a $(d+1)$-dimensional hyperboloid
\begin{align}
  -(X^{-1})^2-(X^0)^2+\sum_{i=1}^d (X^i)^2
  =& -\ell^2
     \label{eq:hyperboloid}
\end{align}
embedded in $(d+2)$-dimensional flat space with signature $(-,-,+,\ldots)$.

Taking
\begin{align}
  X^{-1} =& \ell\cosh\sigma_\gbl \sin t_\gbl,
  &
  X^0 =& \ell\cosh\sigma_\gbl \cos t_\gbl,
  &
    X^i =& \ell\sinh\sigma_\gbl \Omega_\gbl^i,
           \label{eq:hyperboloid_and_global_coordinates}
\end{align}
where $\Omega_\gbl^i$ are coordinates which embed $S^{d-1}$ into $\mathbb{R}^d$. For example, in $d=3$, $\Omega_\gbl^i=(\cos\theta_\gbl,\sin\theta_\gbl\cos\varphi_\gbl,\sin\theta_\gbl\sin\varphi_\gbl)$. Note that we take $t_\gbl\in(-\infty,\infty)$ so that we get a universal cover of the hyperboloid which eliminates closed time-like curves. We shall call $(t_\gbl,\sigma_\gbl,\Omega_\gbl^i)$ global coordinates and, whenever there is potential for confusion, we will use subscript $\gbl$ to distinguish these from other sets of coordinates. The resulting metric is
\begin{align}
  ds^2
  =& \ell^2(-\cosh^2\sigma_\gbl dt_\gbl^2 + d\sigma_\gbl^2 + \sinh^2\sigma_\gbl d\Omega_\gbl^2).
     \label{eq:globalAdS_metric_sigma}
\end{align}
Note that the boundary in these coordinates is at $\sigma_\gbl=+\infty$. 

We can bring the boundary to a finite coordinate $\rho_\gbl=\pi/2$ by taking
\begin{align}
  \tan \rho_\gbl
  =& \sinh\sigma_\gbl \label{globalAdS_rho_sigma}\\
  ds^2
  =& \frac{\ell^2}{\cos^2\rho_\gbl}(- dt_\gbl^2 + d\rho_\gbl^2 + \sin^2\rho_\gbl d\Omega_\gbl^2). \label{eq:globalAdS_metric_rho}
\end{align}
 Sometimes, we will also call $(t_\gbl,\rho_\gbl,\Omega_\gbl^i)$ global coordinates.

\begin{figure}
  \centering
  \includegraphics[width=0.1\textwidth]{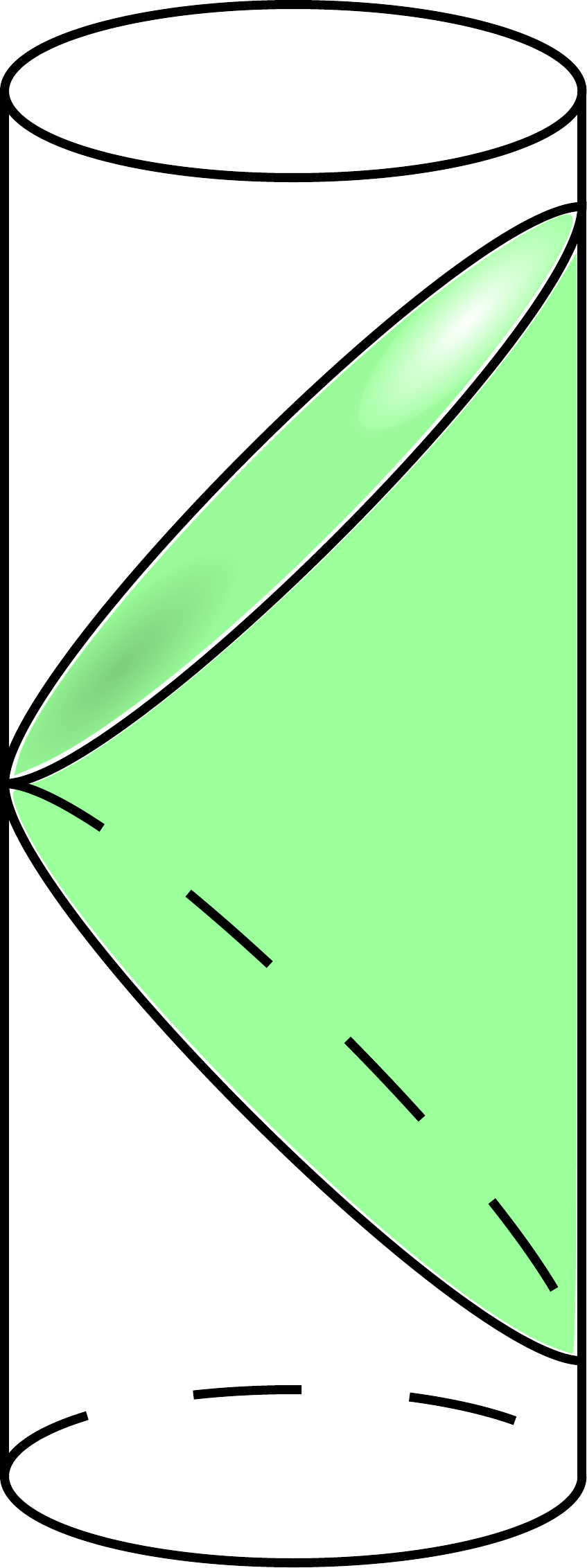}
  \caption{The Poincar\'e coordinates $(t_\Pcr,z_\Pcr,x_\Pcr^i)$ only cover a portion of Lorentzian $\AdS$, called the Poincar\'e patch, shown in green. The entire solid cylinder is covered by the global coordinates $(t_\gbl,\rho_\gbl,\Omega_\gbl^i)$.}
  \label{fig:globalAdS_PoincarePatch}
\end{figure}

To make the boundary geometry Minkowski, take
\begin{align}
  \frac{z_\Pcr}{\ell}
  =& \frac{\ell}{X^0-X^1}
     = \frac{1}{\cosh\sigma_\gbl \cos t_\gbl - \sinh\sigma_\gbl \cos\theta_\gbl}
  \label{eq:Poincare_z}\\
  \frac{t_\Pcr}{z_\Pcr}
  =& \frac{X^{-1}}{\ell}
     = \cosh\sigma_\gbl \sin t_\gbl
  \label{eq:Poincare_t}\\
  \frac{x_\Pcr^i}{z_\Pcr}
  =& \frac{X^i}{\ell}
     = \sinh \sigma_\gbl \Omega_\gbl^i \qquad(i=2,\ldots,d),
     \label{eq:Poincare_x}
\end{align}
where $\Omega_\gbl^1=\cos\theta_\gbl$. We shall call $(t_\Pcr,z_\Pcr,x_\Pcr^i)$ Poincar\'e coordinates, and use subscript $\Pcr$ to refer to these when there is potential for confusion with other coordinates. The metric reads
\begin{align}
  ds^2
  =& \frac{\ell^2}{z_\Pcr^2}[dz_\Pcr^2 -dt_\Pcr^2 + dx_\Pcr^2]
     = \frac{\ell^2}{z_\Pcr^2}[dz_\Pcr^2 +d\tau_\Pcr^2 + dx_\Pcr^2],
     \label{eq:Poincare_metric}
\end{align}
with the latter written in Euclidean time. The Lorentzian Poincar\'e coordinates cover only a part of Lorentzian $\AdS$, called the Poincar\'e patch, shown in Figure \ref{fig:globalAdS_PoincarePatch}. The Euclidean Poincar\'e coordinates cover all of Euclidean global $\AdS_3$.

For $d=2$, let us further define hyperbolic Poincar\'e coordinates,
\begin{align}
  z_\Pcr =& r_\Pcr \cos\theta_\Pcr, & x_\Pcr=& r_\Pcr\sin\theta_\Pcr \cosh\phi_\Pcr, & t=& r_\Pcr\sin\theta_\Pcr\sinh\phi_\Pcr.
  \label{eq:hyperbolic_Poincare_coordinates}
\end{align}
These cover the domain of dependence of the half-space $x_\Pcr>0$ on the $t_\Pcr=0$ slice. The metric reads
\begin{align}
  ds^2
  =& \frac{\ell^2}{r_\Pcr^2\cos^2\theta_\Pcr} \left[
     dr_\Pcr^2 + r_\Pcr^2(d\theta_\Pcr^2 -\sin^2\theta_\Pcr d\phi_\Pcr^2)
     \right].
     \label{eq:hyperbolic_Poincare_metric}
\end{align}
The Euclidean equivalent of hyperbolic Poincar\'e coordinates are just the usual spherical coordinates with hyperbolic functions in \eqref{eq:hyperbolic_Poincare_coordinates} above replaced with their trigonometric counterparts. Spherical Poincar\'e coordinates
\begin{align}
  z_\Pcr =& r_\Pcr \cos\theta_\Pcr,
       \qquad x_\Pcr= r_\Pcr\sin\theta_\Pcr \cos\varphi_\Pcr,
       \qquad \tau_\Pcr= r_\Pcr\sin\theta_\Pcr\sin\varphi_\Pcr \label{eq:spherical_Poincare_coordinates}\\
  ds^2
  =& \frac{\ell^2}{r_\Pcr^2\cos^2\theta_\Pcr} \left[
     dr_\Pcr^2 + r_\Pcr^2(d\theta_\Pcr^2 +\sin^2\theta_\Pcr d\varphi_\Pcr^2)
     \right] \label{eq:spherical_Poincare_metric}
\end{align}
cover all of Euclidean global $\AdS_3$.

\section{Metrics for $\AdS$ black hole}
\label{sec:Sch_FG_metrics}
The Schwarzschild metrics for $(d+1)$-dimensional $\AdS$ black holes take the form
\begin{align}
    ds^2
  =& -f(r_\Sch) dt_\Sch^2 + \frac{dr_\Sch^2}{f(r_\Sch)} + r_\Sch^2 d\Xi_\Sch^2.
     \label{eq:blackhole_Schwarzschild}
\end{align}
The shape of the black hole horizon is described by an integer $k\in\{-1,0,1\}$, with $-1,0,1$ corresponding to hyperbolic, planar, and spherical horizons. Correspondingly, $d\Xi_\Sch^2$ is the squared line element on a $(d-1)$-dimensional hyperboloid, plane, or sphere, scaled as necessary with the $\AdS$ length scale $\ell$ to make dimensions work out in the above. The function $f(r_\Sch)$ is given by
\begin{align*}
    f(r_\Sch)
    =& \frac{r_\Sch^2}{\ell^2} - \frac{\mu}{r_\Sch^{d-2}} + k,
\end{align*}
where $\mu$ parametrizes the mass $M$ of the black hole\footnote{The relation for black hole mass $M$ here follows the conventions of \cite{Witten:1998zw}; an alternative convention further shifts $M$ such that a black hole with zero horizon radius is massless.} in the spherical case $k=1$:
\begin{align*}
  \mu
  =& \frac{16\pi G_N M}{(d-1)\Vol(S^{d-1})}, \qquad(k=1)
\end{align*}
with $G_N$ and $\Vol(S^{d-1})$ respectively being Newton's constant and the volume of a $(d-1)$-sphere.

For the $d=2$ (BTZ) case we are interested in, the metric \eqref{eq:blackhole_Schwarzschild} can be written with
\begin{align}
  f(r_\Sch) =& \frac{r_\Sch^2-r_\hrz^2}{\ell^2},
               \label{eq:blackhole_Schwarzschild_f}
\end{align}
where $r_\hrz$ is the horizon radius. 
In this case, $\Xi_\Sch=\theta_\Sch$ is simply the angular coodinate of a circle with identification
\begin{align}
  \theta_\Sch + 2\pi
  \cong \theta_\Sch.
  \label{eq:BTZ_identification_Schwarzschild}
\end{align}

\subsection{Euclidean BTZ black hole as global $\AdS_3$}
\label{sec:sidewaysGlobalAdS_asBlackHole}
Here, we note that the Euclidean BTZ black hole is equivalent to pure global $\AdS$ upon swapping the roles of spacial and temporal coodinates. Let us define
\begin{align}
  \cos \rho_\Sch \equiv \frac{r_\hrz}{r_\Sch},
  \label{eq:SchwarzschildAdS_rho_r}
\end{align}
then from \eqref{eq:blackhole_Schwarzschild} and \eqref{eq:blackhole_Schwarzschild_f}, we see that the metrics for Lorentzian $(t_\Sch,\rho_\Sch,\theta_\Sch)$ and Euclidean $(\tau_\Sch,\rho_\Sch,\theta_\Sch)$ Schwarzschild coordinates are
\begin{align}
  ds^2
  =& \frac{\ell^2}{\cos^2\rho_\Sch} \left[
     \left(\frac{r_\hrz}{\ell}\right)^2 d\theta_\Sch^2 + d\rho_\Sch^2
     - \left(\frac{r_\hrz}{\ell^2}\right)^2 \sin^2\rho_\Sch dt_\Sch^2
     \right] \label{eq:SchwarzschildAdS_LorentzianMetric} \\
  ds^2 =& \frac{\ell^2}{\cos^2\rho_\Sch} \left[
          \left(\frac{r_\hrz}{\ell}\right)^2 d\theta_\Sch^2 + d\rho_\Sch^2
          + \left(\frac{r_\hrz}{\ell^2}\right)^2 \sin^2\rho_\Sch d\tau_\Sch^2
          \right].
          \label{eq:SchwarzschildAdS_EuclideanMetric}
\end{align}
These are just metrics in the global $\AdS_3$ cylinder, but turned sideways:
\begin{align}
  \tau_\gbl \to& \frac{r_\hrz}{\ell} \theta_\Sch, & \theta_\gbl \to \frac{r_\hrz}{\ell^2} \tau_\Sch.
                            \label{eq:globalAdS_to_SchwarzschildAdS_coords}
\end{align}
From wanting $\tau_\Sch$ to go from $0$ to the inverse temperature $\beta$ as we circle around the cylinder, we deduce the following relationship between the black hole radius and temperature:
\begin{align}
  \frac{2\pi}{\beta} =& \frac{r_\hrz}{\ell^2}.
                        \label{eq:blackHole_temperature_radius}
\end{align}

Analogous to \eqref{globalAdS_rho_sigma} and \eqref{eq:globalAdS_metric_sigma}, we shall write
\begin{align}
  \sinh \sigma_\Sch =& \tan \rho_\Sch \label{eq:SchwarzschildAdS_rho_sigma}\\
  ds^2
  =& \ell^2\left[
     \left(\frac{r_+}{\ell}\right)^2 \cosh^2\sigma_\Sch d\theta_\Sch^2
     + d\sigma_\Sch^2
     - \left(\frac{r_+}{\ell^2}\right)^2 \sinh^2\sigma_\Sch dt_\Sch^2
     \right].
     \notag
\end{align}
which is useful since lines of constant $\theta_\Sch,t_\Sch$ have proper length measured by $\ell d\sigma_\Sch$.

\section{Special Functions}
Most of this is taken from \cite{abramowitz+stegun} and \cite{NIST:DLMF}.

\subsection{Hypergeometric function}
\label{sec:hyperF}
The hypergeometric differential equation is a second order complex differential equation
\begin{align}
  z(1-z) w'' + [c-(a+b+1)z] w' - ab w
  =& 0 \label{eq:hyperDE}
\end{align}
containing three regular singular points: $0,1,\infty$. Around each regular singular point are two linearly independent solutions, usually given by
\begin{align}
  \text{$z$ around $0$}:&
  & &\hyperF(a,b;c;z), \label{eq:hyperDE_z0_sol1}\\
                        &
  & &z^{1-c} \hyperF(1+a-c,1+b-c;2-c;z), \label{eq:hyperDE_z0_sol2}\\
  \text{$z$ around $1$}:&
  & &\hyperF(a,b;1+a+b-c;1-z), \\
                        &
  & &(1-z)^{c-a-b} \hyperF(c-a,c-b;1+c-a-b;1-z), \\
  \text{$z$ around $\infty$}:&
  & & z^{-a}\hyperF(a,1+a-c;1+a-b;z^{-1}), \\
                        &
  & & z^{-b}\hyperF(b,1+b-c;1+b-a;z^{-1}), \label{eq:hyperDE_zinfty_sol2}
\end{align}
where the hypergeometric function $\hyperF$ is defined by
\begin{align}
  \hyperF(a,b;c;z)
  \equiv& \sum_{n=0}^\infty \frac{(a)_n(b)_n}{(c)_n} \frac{z^n}{n!},
          \label{eq:hyperF}
\end{align}
where $(q)_n$ is the Pochhammer symbol,
\begin{align*}
  (q)_n
  \equiv& \prod_{j=0}^{n-1} (q+j).
\end{align*}
The series \eqref{eq:hyperF} converges if $c$ is not a non-positive integer and either $|z|<1$ or both $|z|=1$ and $\Re(c-a-b)>0$. But, the series can be analytically continued elsewhere, provided $z=1,\infty$ are avoided. There are various conditions attached to the solutions \eqref{eq:hyperDE_z0_sol1}-\eqref{eq:hyperDE_zinfty_sol2}. For example, if $c\le 0$ is an integer, then \eqref{eq:hyperDE_z0_sol1} does not exist; if $c\ge 2$ is an integer, then \eqref{eq:hyperDE_z0_sol2} does not exist; if $c=1$, then \eqref{eq:hyperDE_z0_sol1} and \eqref{eq:hyperDE_z0_sol2} are equal. In any of those cases, another, more complicated, solution must be written.

It is also possible to express the solutions \eqref{eq:hyperDE_z0_sol1}-\eqref{eq:hyperDE_zinfty_sol2} in alternative forms using
\begin{align}
  \hyperF(a,b;c;z)
  =& (1-z)^{-a} \hyperF\left(a,c-b;c;\frac{z}{z-1}\right) \label{eq:hyperF_equivalent_form1}\\
  =& (1-z)^{-b} \hyperF\left(c-a,b;c;\frac{z}{z-1}\right) \label{eq:hyperF_equivalent_form2}\\
  =& (1-z)^{c-a-b} \hyperF(c-a,c-b;c;z). \label{eq:hyperF_equivalent_form3}
\end{align}

For $|\arg(1-z)|<\pi$, we have
\begin{align}
  \hyperF(a,b;c;z)
  =& \frac{\Gamma(c)\Gamma(c-a-b)}{\Gamma(c-a)\Gamma(c-b)} \;\hyperF(a,b;a+b+1-c;1-z) \notag\\
   &+ \frac{\Gamma(c)\Gamma(a+b-c)}{\Gamma(a)\Gamma(b)} (1-z)^{c-a-b} \hyperF(c-a,c-b;1+c-a-b;1-z).
     \label{eq:hyperF_linear_combination1}
\end{align}
Additionally, for $|\arg(z)|,|\arg(1-z)|<\pi$, we have
\begin{align}
  \MoveEqLeft[3]\hyperF(a,b;c;z) \notag\\
  =& \frac{\Gamma(c)\Gamma(c-a-b)}{\Gamma(c-a)\Gamma(c-b)}
     z^{-a} \hyperF(a,a-c+1;a+b-c+1;1-z^{-1}) \notag\\
   &+ \frac{\Gamma(c)\Gamma(a+b-c)}{\Gamma(a)\Gamma(b)}
     (1-z)^{c-a-b} z^{a-c} \hyperF(c-a,1-a;c-a-b+1;1-z^{-1})
     \notag \\
  =& \frac{\Gamma(c)\Gamma(c-a-b)}{\Gamma(c-a)\Gamma(c-b)}
     z^{1-c} \hyperF(1+b-c,1+a-c;a+b-c+1;1-z) \notag\\
   &+ \frac{\Gamma(c)\Gamma(a+b-c)}{\Gamma(a)\Gamma(b)}
     (1-z)^{c-a-b} z^{1-c} \hyperF(1-b,1-a;c-a-b+1;1-z)
     \label{eq:hyperF_linear_combination3}
\end{align}
The last equality was obtained by applying \eqref{eq:hyperF_equivalent_form2} to the previous line.

As $c$ approaches a non-positive integer $-m$, we get
\begin{align}
  \lim_{c\to-m} \frac{\hyperF(a,b;c;z)}{\Gamma(c)}
  =& \frac{(a)_{m+1}(b)_{m+1}}{(m+1)!} z^{m+1} \hyperF(a+m+1,b+m+1;m+2;z).
     \label{eq:hyperF_c_to_nonpositive_integer}
\end{align}
It is obvious from the definition \eqref{eq:hyperF} that, at $z=0$,
\begin{align*}
  \hyperF(a,b;c;0)
  =& 1.
\end{align*}
The value at $z=1$,
\begin{align}
  \hyperF(a,b;c;1)
  =& \frac{\Gamma(c)\Gamma(c-a-b)}{\Gamma(c-a)\Gamma(c-b)},
     \label{eq:hyperF_z1}
\end{align}
can be deduced from \eqref{eq:hyperF_linear_combination1}.

\subsection{Associated Legendre and Ferrers Functions}
\label{sec:LegendreFerrersP}
The associated Legendre function (of the first kind) of degree $a$ and order $b$ is related to hypergeometric functions by
\begin{align*}
  P^b_a(z)
  =& \left(\frac{z+1}{z-1}\right)^{b/2} \frac{\hyperF\left(-a,a+1; 1-b; \frac{1-x}{2}\right)}{\Gamma(1-b)},
\end{align*}
which has a branch cut for $z\in(-1,1)$.
There are identities relating positive and negative degrees and orders,
\begin{align}
  P_{-1-a}^b(z)
  =& P_a^b(z)
     \label{eq:LegendreP_degreeSignFlip} \\
  P_a^{-b}(z)
  =& \frac{\Gamma(a-b+1)}{\Gamma(a+b+1)}\left[
     P_a^b(z)-\frac{2}{\pi} e^{-i\pi b} \sin(\pi b) Q_a^b(z)
     \right],
     \label{eq:LegendreP_orderSignFlip}
\end{align}
where $Q_a^b(z)$ is the Legendre function of the second kind:
\begin{align*}
  Q_a^b(z)
  =& e^{i\pi b} 2^{-a-1} \sqrt{\pi} \frac{\Gamma(a+b+1)}{\Gamma\left(a+\frac{3}{2}\right)} \\
     &\times z^{-a-b-1}(z^2-1)^{b/2} \hyperF\left(
     1+\frac{a+b}{2},
     \frac{1+a+b}{2};
     \nu+\frac{3}{2};
     \frac{1}{z^2}
     \right).
\end{align*}

Along the branch cut, $x\in(-1,1)$, one typically defines the Ferrers function
\begin{align*}
  P^b_a(x)
  =& \frac{1}{2}\left[e^{i\pi b/2} P^b_a(x+i0)+e^{-i\pi b/2} P^b_a(x-i0)\right]
  \\
  =& \left(\frac{1+x}{1-x}\right)^{b/2} \frac{\hyperF\left(-a,a+1; 1-b; \frac{1-x}{2}\right)}{\Gamma(1-b)}.
\end{align*}
For $x\in(0,1)$, we have
\begin{align}
  \hyperF\left(a,b;a+b+\frac{1}{2};x\right)
  =& 2^{a+b-\frac{1}{2}} \Gamma\left(\frac{1}{2}+a+b\right)
     x^{\frac{1}{2}\left(\frac{1}{2}-a-b\right)}
     P^{\frac{1}{2}-a-b}_{a-b-\frac{1}{2}}\left[(1-x)^{1/2}\right].
     \label{eq:hyperF_FerrersP}
\end{align}
At $x=0$, we have
\begin{align}
  P_a^b(0)
  =& \frac{2^b \cos\left[\frac{\pi}{2}(a+b)\right]\Gamma\left(\frac{1+a+b}{2}\right)}{\sqrt{\pi}\Gamma\left(1+\frac{a-b}{2}\right)}
  = \frac{2^b \sqrt{\pi}}{\Gamma\left(\frac{1-a-b}{2}\right)\Gamma\left(1+\frac{a-b}{2}\right)}.
     \label{eq:FerrersP_x0}
\end{align}

\section{Generalized Eigenvalue Problem}
\label{sec:generalized_eigenvalue_problem}
Here, we record some notes on the generalized eigenvalue problem: given Hermitian matrices $A,B$, 
find matrices $V,D$, with $D$ diagonal such that
\begin{align}
  AV
  =& BVD.
     \label{eq:generalized_eigenval_problem}
\end{align}
Here, the columns of $V$ are `generalized eigenvectors' corresponding to `generalized eigenvalues' given by the diagonal entries of $D$. Note that the $B=\mathds{1}$ case reduces to a standard eigenvalue problem.

Let us write the diagonalization of $B$ as
\begin{align*}
  B =& V_B D_B V_B^\dagger.
\end{align*}
Additionally, let
\begin{align}
  A' \equiv& (V_B D_B^{-1/2})^\dagger A V_B D_B^{-1/2},
             \label{eq:Aprime_definition}
\end{align}
be diagonalized as
\begin{align}
  A' =& V_{A'} D_{A'} V_{A'}^\dagger.
        \label{eq:Aprime_diagonalization}
\end{align}
Then, the generalized eigenvalue problem \eqref{eq:generalized_eigenval_problem} is solved by
\begin{align*}
  V=& V_B D_B^{-1/2} V_{A'},
  &
    D =& D_{A'}.
\end{align*}
Substituting these into \eqref{eq:generalized_eigenval_problem} gives
\begin{align*}
  \text{\eqref{eq:generalized_eigenval_problem} LHS}
  =& AV_B D_B^{-1/2} V_{A'},
  & \text{\eqref{eq:generalized_eigenval_problem} RHS}
    =& BV_B D_B^{-1/2} V_{A'} D_{A'}
       = V_B D_B^{1/2} V_{A'} D_{A'},
\end{align*}
which are equal by the definition \eqref{eq:Aprime_definition} and diagonalization \eqref{eq:Aprime_diagonalization} of $A'$.

\bibliographystyle{JHEP}
\bibliography{references}








\end{document}